\documentclass[article,12pt]{elsarticle}
\usepackage[titletoc]{appendix}
\usepackage{ucs}
\usepackage[utf8x]{inputenc}
\usepackage[scaled=1]{couriers}
\usepackage[T1]{fontenc}
\usepackage[english,spanish]{babel}
\usepackage{amsmath, amssymb, amsthm, amsfonts}
\usepackage{textcomp}
\usepackage{color}
\usepackage{calc}
\usepackage{longtable}
\usepackage{lscape}
\usepackage{supertabular}
\usepackage{gensymb}
\usepackage{MnSymbol} 
\usepackage{lscape}
\usepackage{wasysym}
\usepackage{hyperref}
\usepackage{tabulary}
\usepackage{array}
\usepackage{hhline}
\usepackage{ulem}
\usepackage{lineno}
\usepackage{graphics}
\usepackage{natbib}
\usepackage{hypernat}
\usepackage{multicol}

\usepackage{tikz}
\usepackage{lipsum}
\usepackage[T1]{fontenc}
\usepackage{scrextend}

\newtheorem{theorem}{Theorem}

\hypersetup{colorlinks=true, linkcolor=blue, filecolor=blue, urlcolor=blue}

\definecolor{LOgreen}{RGB}{35.445, 165.2, 0}

\newcommand{\comillas}[1]{\textquotedblleft{}{#1}\textquotedblright{}}
\newcommand{\scomillas}[1]{\textquoteleft{#1}\textquoteright{}}
\newcommand{\then}{\text{\wasytherefore{\quad }}}

\newcommand{\E}{\boldsymbol{E}}
\newcommand{\R}{\mathbb{R}}

\newcommand{\Co}{\mathfrak{C}}
\newtheorem{corollary}{Corollary}

\makeatletter

\newcommand\ps@Standard{%
\renewcommand\@oddhead{}%
\renewcommand\@evenhead{}%
\renewcommand\@oddfoot{\thepage}%
\renewcommand\@evenfoot{\@oddfoot}%
\setlength\paperwidth{8.5in}
\setlength\paperheight{11in}
\setlength\voffset{-1in}
\setlength\hoffset{-1in}
\setlength\topmargin{0.7874in}
\setlength\headheight{12pt}
\setlength\headsep{0cm}
\setlength\footskip{12pt+0.1965in}
\setlength\textheight{11in-0.7874in-0.7874in-0cm-12pt-0.1965in-12pt}
\setlength\oddsidemargin{0.7874in}
\setlength\textwidth{8.5in-0.7874in-0.7874in}
\renewcommand\thepage{\arabic{page}}
\setlength{\skip\footins}{0.0398in}
\renewcommand\footnoterule{\vspace*{-0.0071in}
\noindent\textcolor{black}{\rule{0.25\columnwidth}{0.0071in}}
\vspace*{0.0398in}}
}
\makeatother
\journal{{ }}
\begin{document}
\selectlanguage{english}



\citestyle{numbers} 
\citestyle{square} 
\citestyle{citesep={,}} 
\citestyle{sort}
\citestyle{compress} 
\bibliographystyle{unsrtnat}   

\begin{frontmatter}
	\title{Hidden  chaos factors inducing random walks which reduce hospital operative efficiency.}
	
	\author[lf]{A.J. Rodr\'{\i}guez-Hern\'{a}ndez MD, MSc}
	\author[sc]{C. Sevcik, MD, PhD, Professor Emeritus\corref{cor}}
	
	\address[lf]{Medical Director, Hospital La Fuenfr\'{\i}a. , Carretera de las Dehesas, CercedillaPostal Code 28470,Madrid, Spain.}
	
	\address[sc]{Centro de Biof\'{\i}sica y Bioqu\'{\i}mica, Instituto Venezolano de Investigaciones Cient\'{\i}ficas, Caracas, Venezuela}

	\cortext[cor]{C. Sevcik , SciMeDAn, Av. Paral\textperiodcentered{}lel 124, Ent. 2B, Postal Code 08015, Barcelona, Spain. Mobile:+34 697 66 84 0. email: carlos.sevcik.s@gmail.com. ORCID: 0000-0003-3783-6541}

	\begin{abstract}
		La Fuenfr\'{\i}a Hospital (LFH) operative parameters such as: hospitalised patients; daily admissions and discharges were studies for the hospital as a whole, and per each Hospital\textquoteright{s} service unit (just called \scomillas{service} here). Data were used to build operative parameter value series and their variations. Conventional statistical analyses and fractal dimension analyses were performed on the series.  Statistical analyses indicated that the data did not follow a Gauss (i.e. \scomillas{normal}) distribution, thus nonparametric statistical analyses were chosen to describe data.  The sequence of daily admitted admissions and patients staying on each service were found to be a kind of random series of a kind called \textit{random walks} (\textit{Rw}). \textit{Rw}, sequences where what happens next, depends on what happens now plus a random variable. \textit{Rw} analysed with parametric or nonparametric statistics may simulate cycles and drifts which resemble seasonal variations or fake trends which reduce the Hospital\textquoteright{s} efficiency. Globally, inpatients \textit{Rw}s in LFH, were found to be determined by the time elapsed between daily discharges and admissions. The factors determining LFH \textit{R} were found to be the difference between daily admissions and discharges.  The analysis suggests discharges are replaced by admissions with some random delay and that the random difference determines LFH \textit{Rw}s. The daily difference between hospitalised patients follows the same statistical distribution as the daily difference between admissions and discharges. These suggest that if the daily difference between admissions and discharges is minimised, i.e., a patient is admitted without delay when another is discharged, the number of inpatients would fluctuate less and the number of unoccupied  beds would be reduced optimising the Hospital service.
	\end{abstract}


\end{frontmatter}

\section{Introduction.}\label{S:Preamble}

The efficiency of a service providing institution depends largely on avoiding fluctuations between active and inactive periods. This is true for hotels, airplanes companies, hospitals or any manufacturing production lines. Control and predictability rationalise costs and inventory expenses, this has long been known to proponents of total quality control management \citep{Deming1986}. 

Hospitals in general, are service institutions which revive a random queue of service demanders (patients) whose admissions depends both on patient\textquoteright{s}  availability and on hospital\textquoteright{s} capacity to attend or admit a patient. In a hospital operating at full capacity a patient must be discharged prior to the admission of a new patient. If a hospital is a part of a network of hospitals, the situation becomes even more complex, since it depends on hospital\textquoteright{s}communication between the discharge and admission and discharge sections/systems, and been hospitals in the network which transfer patients to another hospital. In the latter situation transportation between hospitals may be another factor depending on availability of a transport mean, or if this depends on the patient\textquoteright{s} on whether those resources are available, immediately or with some sort of delay. All this factors conform a complex set which determine hospital efficacy, and introduce complex non linearities in the system, and yet this non linearities, to our knowledge, are never considered when hospital efficiency is evaluated. 

Non linear systems such as described in the preceding paragraph, are poor candidates to analyse with standard statistical means. This has been shown for may events such as earthquakes, wetter, adult and foetal hearth beat, epidemics, highway traffic, stock market fluctuations, machine failures and many more \cite{Lorenz1963, Mandelbrot1963, Mandelbrot1967, Mandelbrot1983, Meltzer1991, Pincus1991, Bak1995, Telesca2005, Mitchell2020}. Still, the common practice to evaluate an detect institutional efficiency (or lack of it) is generally limited to actuarial studies and some descriptive statistics loaded into administrative data bases. Budgets, personnel hiring, firing, promotions, demotions, and even hospital closures, may result based on these questionable analyses. This paper describes the analysis of of a real hospital, and benefits from the development of a simple algorithm \cite{Sevcik1998a} to estimate  the Hausdorff-Besicovitch \cite{Besicovitch1929, Hausdorff1918} dimension using a a variety of box-counting algorithm, the so estimated fractal dimension has been called with increasing frequency: \comillas{Sevcik\textquoteright{s} dimension}\cite{Sharma2013, Diao2017, Shi2018, Nepiklonov2020, Xue2020,  Kolodziej2020}. Thus,  from paper on, we will start calling \comillas{Sevcik\textquoteright{}s fractal dimension}: $ D_s $ since this actually avoids some formal mathematical confusions. GogSch\footnote{\textcopyright{}2018 Google LLC All rights reserved. GogSch and the Google logo are registered trademarks of Google LLC.} \cite{GoogleScholar2020}  lists 152 cases of fractal analysis, in many fields, up to November 9, 2020 \cite{Sevcik2020} with this \cite{Sevcik1998a} algorithm.

La Fuenfr\'{\i}a Hospital  (LFH, Servicio Madrile\~{n}o de Salud, SE\textit{R}MAS) is a public mid and long term stay hospital [in the Madrid\textquoteright{s} (Spain) Autonomous Region] for chronic patients. Spain has a life expectancy of 83.61 years (the 6\textsuperscript{th} longest in the world), and is growing at a  $ 0.21 $0\% per year rate) \cite{Macrotrends2019}. This fast growth makes Spain a candidate to become the country with world\textquoteright{}s longest life expectancy in the near future.

In 2019, Spain had a population of 46,738,578 inhabitants, and 26.5\% ($ \approx 12.2 $ million)) are above 60 years of age (calculated based on \citep{Mathers2015, SpainPop2019, Macrotrends2019, LifeExp2020}). One would then expect that hospitals for chronic diseases would have a significant and constant demand. Yet when La Fuenfr\'{\i}a Hospital curve of patients staying al the Hospital (abbreviated \textbf{InP}) during the period under (Figure \ref{F:DailyHospitalised}A) LFH fluctuated between being completely full followed by periods with only $ \approx 52 $\% occupancy (the hospital has 192 beds). Analysed with classical administrative and statistical tools, this behaviour could be attributed to some kind of service demand seasonal periodicity or, even worse, attributed to administrative deficiency. The fact remains, that no matter what explanation is given, this behaviour is undesirable, reduces hospital efficiency, increases costs, and thus, the cause must be identified and eliminated as completely as possible.

Here we use fractal analysis, Fourier transforms,spectral analysis, and conventional statistics, to show that small hidden factors may have indeed a dramatic impact to reduce hospital\textquoteright{s} operative efficiency. the purpose of the paper is to show that factors not usually considered in hospital performance analysis may have a huge impact on hospital efficacy.

\section{Methods.}

\subsection{Statistical methods and spectral density (SD) calculation.}

\subsubsection{Statistical methods.}

Medians and their 95\% confidence interval (CI) were calculated using the nonparametric Hodges and Lehman method \cite{Hollander1973}. Random distribution about the median ($ \widehat{y} $) was analysed with the above and below \textit{runs} (one or more data appearing as a sequence above or below the median) methods of Mood \cite{Mood1940} and Wilks \cite{Wilks1962, Guttman1971}.  Gaussianity (also known as normality) was verified with the  Jarque-Bera \cite{Bera1981a}, then robustified Jarque-Bera-Gel \cite{Gel2008}  and the Shapiro Wilk \cite{Shapiro1965} tests. Linear regressions were carried out using the Theil nonparametric method \cite{Hollander1973} and correlations estimated with the Spearman rank correlation coefficient \cite{Hollander1973}. Sequences were compared with the nonparametric Smirnov \cite{Smirnov1939} test based on Kolmogorov statistics \cite{Kolmogorov1933}. The procedures to generate random sequences to verify the differences between their fractal dimension were described elsewhere  \cite{Sevcik1998a, DSuze2010a, DSuze2015a, Sevcik2018}

\subsubsection{Fast Fourier Transforms (FFT) and calculating spectral density.}

FFT were calculated with a C++ (using g++ version 5.4.0 20160609 with C++14 standards, ww.gnu.org) implemented Cooley and Tukey algorithm \cite{Cooley1965}, with Hamm windowed data \cite{Blackman1959}. Spectral densities were then obtained from the FFT real and imaginary components \cite{Blackman1959}.
\begin{figure}[h!]
	\begin{center}
		\includegraphics[width=12cm]{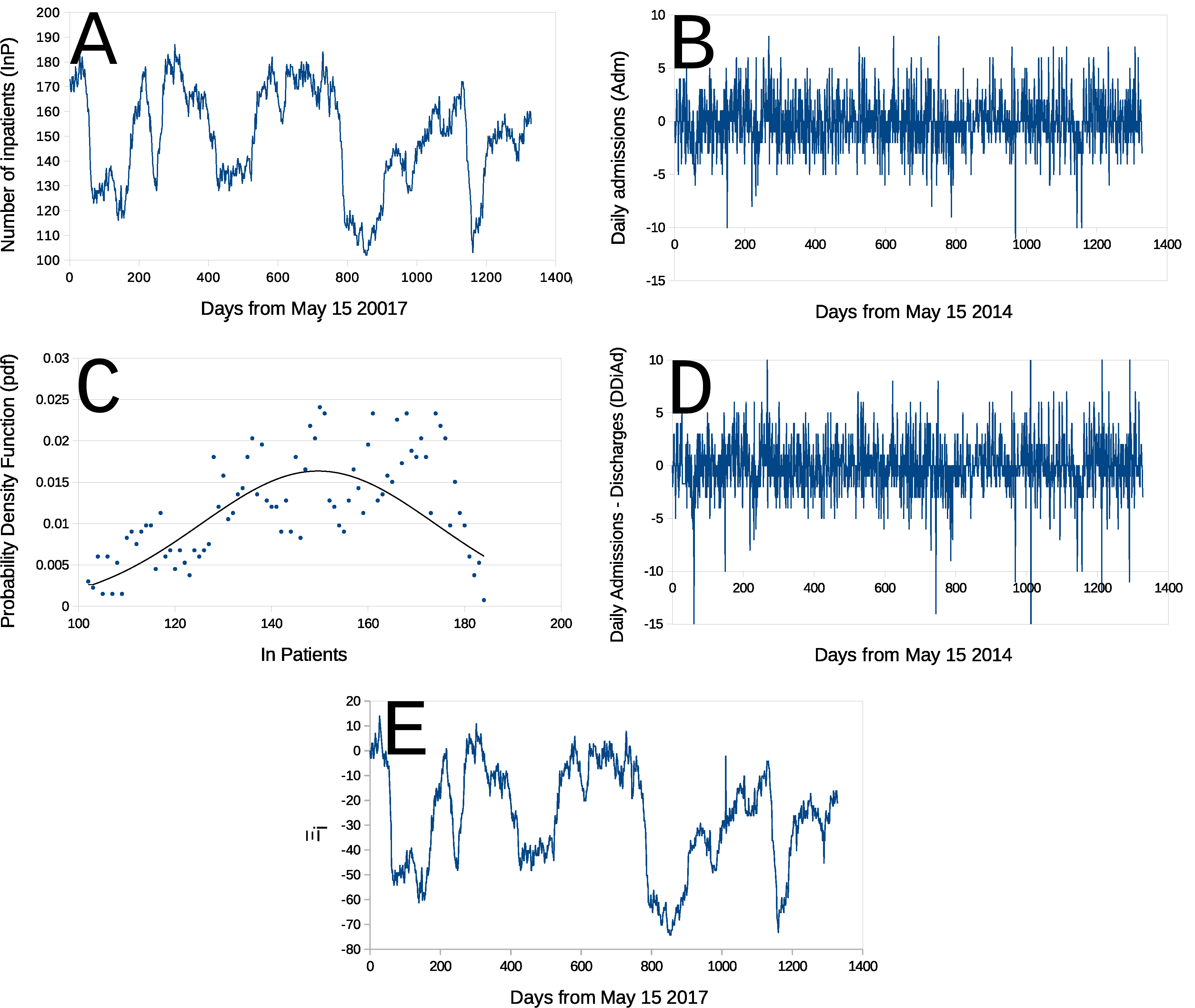}
		\caption{\textbf{Patients staying every day (\textbf{InP}) at La  Fuenfr\'{\i}a Hospital from May 1 2014 and December  19 2017.} Panels are: \textbf{Panel A}- Number of inpatients (\textbf{InP}) at La Fuenfr\'{\i}a Hospital  per day during the study duration; \textbf{Panel B}- Daily variations of inpatients (\textbf{DInP}) at the Hospital [$ \Delta y $ in Eq. (\ref{E:SecDifGen})]; \textbf{Panel C}-  Hospital\textquoteright{}s daily stays (Panel A) \textbf{p}robability \textbf{d}ensity \textbf{f}unction (\textit{pdf}, the probability of observing a given mensurable or enumerable random event) for any number of inpatients (\textbf{InP}) staying at the Hospital observed during the study period ($ \bullet $) \cite{Wilks1962}, a Gauss line ($ N[\bar{y}, s[y] $) with the same mean and variance ($ s[y]^2 $) as the set of 1329 data points making the series in Panel A ($ \bar{y}=149.2 $, $ s[y]=24.4 $); \textbf{Panel D}- Daily difference between admissions  (Figure 2A) and discharges (Figure 2C) at the Hospital (\textbf{DDiAd}). \textbf{Panel E}- Random walk $ \boldsymbol{\Xi_i} $ built for \textbf{DDiAd} (Panel D) using Eq. (\ref{E:DDiAd}), please note the extreme resemblance to the sequence of \textbf{InP} in Panel A. In all cases abscissa is number of days after May 1 2014. The largest negative or positive peaks are off scale outliers.}\label{F:DailyHospitalised}
	\end{center}
\end{figure}

\subsection{Building the time series.}\label{S:Admissions_Discharges}

\begin{figure}[h!]
	\begin{center}
		\includegraphics[width=12cm]{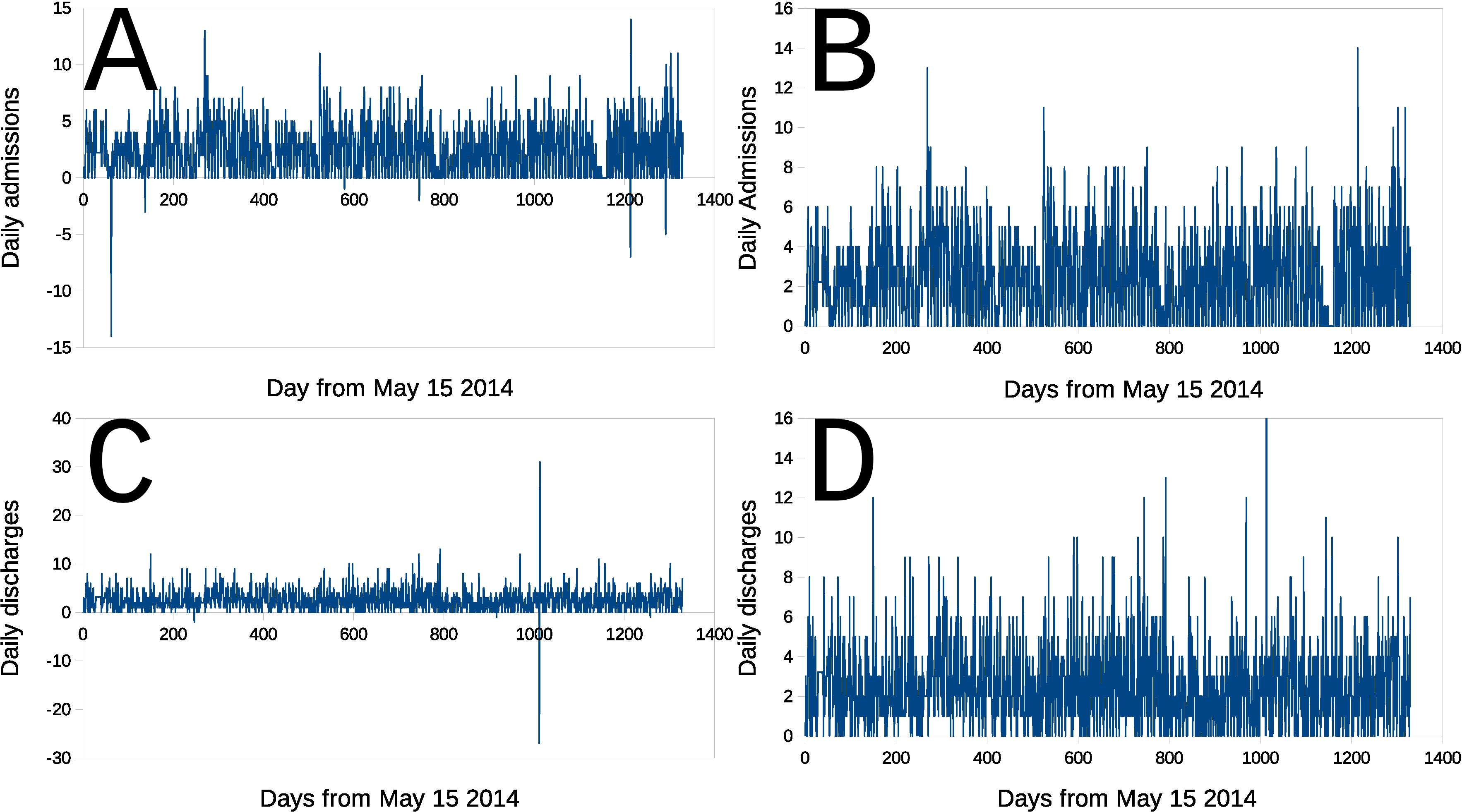}
		\caption{\textbf{Sequences of hospitalised and daily discharges from La Fuenfr\'{\i}a Hospital.} Panels are: \textbf{A}- Daily admissions (\textbf{Adm}) sequence; \textbf{B}- Same as panel A, but scale modified to show only data $ \geqslant0 $; \textbf{C}- Daily discharge (\textbf{Dis}) sequence; \textbf{D}- Same as panel C,  but scale modified to show only data $ \geqslant0 $. In all cases abscissa is days elapsed since May 1 2014, Ordinate is number of daily admissions or discharges determined as indicated in Secretion \ref{S:Admissions_Discharges}. Negative peak represent transcription errors of data made available for this study, perhaps positive outliers could have the same origin, but this is harder to demonstrate.}\label{F:Admissions_Discharges}
	\end{center}
\end{figure}

Series were built for La Fuenfr\'{\i}a Hospital\textquoteright{}s admissions,  discharges (\textbf{Dis}) and hospitalised patients, called \comillas{inpatients}, and shortened \textbf{InP} in this paper. The Hospital depends on Autonomous Madrid Region area Public Health System Network [Servicio Madrileño de Salud (SERMAS)] which, like most Spain, requires that all hospital information in the area is lodged into a central data base. In mid term stay hospital this information is kept using a program called SELENE\textsuperscript{\textcopyright}. {SELENE\textsuperscript{\textcopyright}} is jointly developed by UTE Siemens-- INDRA \cite{SELENE2007}. {SELENE\textsuperscript{\textcopyright}} keeps data in a central database for SERMAS. Data were retrieved from {SELENE\textsuperscript{\textcopyright}} as monthly spreadsheets for each kind of information, and were integrated in a single file covering our entire period of study for each kind of information. 

To build the sequences for this study, we used data collected monthly as {SELENE\textsuperscript{\textcopyright} Hospital\textquoteright{s} spreadsheet files called \comillas{\textit{Tablas de Mando}} which we translate\textit{ Command panels}. In them, information appears as daily data for La Fuenfr\'{\i}a Hospital from   5/1/2014 up to 12/31/2014, then they run from  January 1 to December 31 for years 2015 and 2016, and from 1/1/2017 up to 12/19/2017. 
	
	\begin{table}[h!]
		\begin{center}
			\caption{Fractal dimension $ D_s $ estimated for La Fuenfr\'{\i}a Hospital sequences. }\label{T:FracDimSer}
			\begin{tabular}{cccc}
				\hline \hline
				{Sequence} & $ \qquad \qquad $&
				$ D_s $ &
				$ \sqrt{ \text{var}[D_s]}$ \\
				\hline
				Admissions (\textbf{Adm}  & &
				$ 1.49725 $  &
				$ 0.00356 $ \\
				Discharges (\textbf{Dis} $ D_s $) && 
				$ 1.41904 $ &
				$ 0.00400 $ \\
				In Patients (\textbf{InP} &&
				$ 1.33806 $ &
				$ 0.00320 $ \\
				InP variation (\textbf{DInP}  & &
				$ 1.56432 $ &
				$ 0.00302 $ \\
				Adm-Dis (\textbf{DDiAd})  & &
				$ 1.43369 $ &
				$ 0.00401 $ \\
				$ \boldsymbol{\Xi_i} $ \textit{Rw}  &&
				$ 1.33922 $ &
				$ 0.00382$\\
				Brownian noise\textbf{\textsuperscript{\S}}& &
				$ 1.28855 $&
				$ 0.03837 $ \\
				White noise\textbf{\textsuperscript{\S}} &&
				$ 1.59825 $ &
				$ 0.01039 $\\
				\hline \hline
			\end{tabular} 
			\begin{center}
				\textbf{Dimension} $ \boldsymbol{D_s} $ calculated with Eq.  (\ref{E:D_h}); $ s(d)=\sqrt{\boldsymbol{\text{var}(D_s)}} $  calculated with Eq, (\ref{E:varD}), for all sequences $ N=1329 $. \textbf{Labels mean}: \textbf{InP},in patients at  La Fuenfr\'{\i}a Hospital (Fig. \ref{F:DailyHospitalised}A); \textbf{Adm}, admissions to La Fuenfr\'{\i}a Hospital ( Fig. \ref{F:Admissions_Discharges}A);\textbf{Dis}, discharges from La Fuenfr\'{\i}a Hospital (Fig. \ref{F:Admissions_Discharges}D) ;  \textbf{InP}, daily sequence of inpatients (Fig. \ref{F:DailyHospitalised}A); \textbf{DInP}, InP daily variation ( Fig. \ref{F:DailyHospitalised}B)  ; \textbf{DDiAd}, Adm-Dis (Fig. \ref{F:DailyHospitalised}D); $ \boldsymbol{\Xi_i} $, \textit{Rw} built for \textbf{DDiAd} curve in Fig. \ref{F:DailyHospitalised}E using Eq. (\ref{E:DDiAd}); \textbf{Brownian noise}, Brownian noise calculated as indicated in Sub-Section \ref{S:RandFract} and Eq. (\ref{E:BroGaus}). \textbf{White noise}, White noise calculated as indicated in Sub-subsection \ref{S:RandFract}.  \textbf{{\S}} superscript, indicates that $ D_s $ is the average of  $ M=100 $ sequences with $N= 1329 $ evaluated with Eq. (9), total variance was  calculated with Eq. (\ref{E:varTotal}). .
			\end{center}
		\end{center}
	\end{table}
	
	Data was routinely hospital\textquoteright{s} statistics, was not recorded purportedly for this study, and we found transcription errors in the SELENE\textsuperscript{\textcopyright}-exported data base, which were dealt with as follows:
	\begin{itemize}
		\item {January 1\textsuperscript{st} data corrections}. An obvious error detected in the time series was that more than 900 admissions or discharges appeared for 1/1/2015 and 1/1/2016, an impossibility for a hospital with 192 beds. The source of such errors appears to be a bug in the software used to gather data which did not  evaluate the first datum of the year correctly. Those points were replaced by averaging December 31 of the previous year with January 2 of the current year, respectively. This error affected 2 out of 1329 days, or $ \approx 0.16 $\% of data.
		\item {\textbf{Missing data corrections}. In Admissions (\textbf{Adm}) and Discharges (\textbf{Dis}) series we found a gap without information from 5/29/2014 to 6/6/2014 (both inclusive), in daily  in--patient . This gap ( 9 out of 1329 days, or $ \approx 0.67 $\% of data) was closed filling the \textit{data on daily changes in the sequences for the gap days }with a same value calculated as
			\begin{equation}\label{E:GapCor}
				\left\lbrace y_{(\text{6/6/2014})} \leftrightarrow y_{(\text{5/29/2014})} 
				\right\rbrace = \tfrac{y_{(\text{6/7/2014)}}-y_{\text{(5/28/2014)}}}{9},
			\end{equation}
			where dates follow the \comillas{mm/dd/yyyy} style, and the double headed arrow at left of Eq. (\ref{E:GapCor}) points at the limits of the gap to fill. In Eq. (\ref{E:GapCor}), as well as elsewhere in this paper, brackets are used to indicate sets.}
		\item Daily differences between \textbf{Adm}, \textbf{Dis} and \textbf{InP}.
		Once daily data admissions (\textbf{DAdm}), daily discharges (\textbf{DDis}) and daily inpatients (\textbf{DInP}) sequences were built, new series were developed by subtracting to each daily value, the immediately preceding day\textquoteright{s} similar datum. A set of sequences of daily differences
		\begin{equation}\label{E:SecDifGen}
			\Delta y_{t}   =   \begin{cases}
				0 &\implies t=0\\
				y_{t} - y_{t-\Delta t} &\implies t > 0
			\end{cases}
		\end{equation}
		with $ t =0,1, \ldots, 1328 $ days, was built. In Eq. (\ref{E:SecDifGen}) each $ t $ is a given day, and  $ \Delta t  =1$ is a one day time difference, the first day was arbitrarily taken as 0.  The procedure was repeated for admissions, discharges and stays to obtain the corresponding daily differences sequences for the three kinds of parameters shown in Figures \ref{F:Admissions_Discharges}A and \ref{F:Admissions_Discharges}B.
	\end{itemize}

\section{Results.}
\begin{figure}[h!]
	\begin{center}
		\includegraphics[width=12cm]{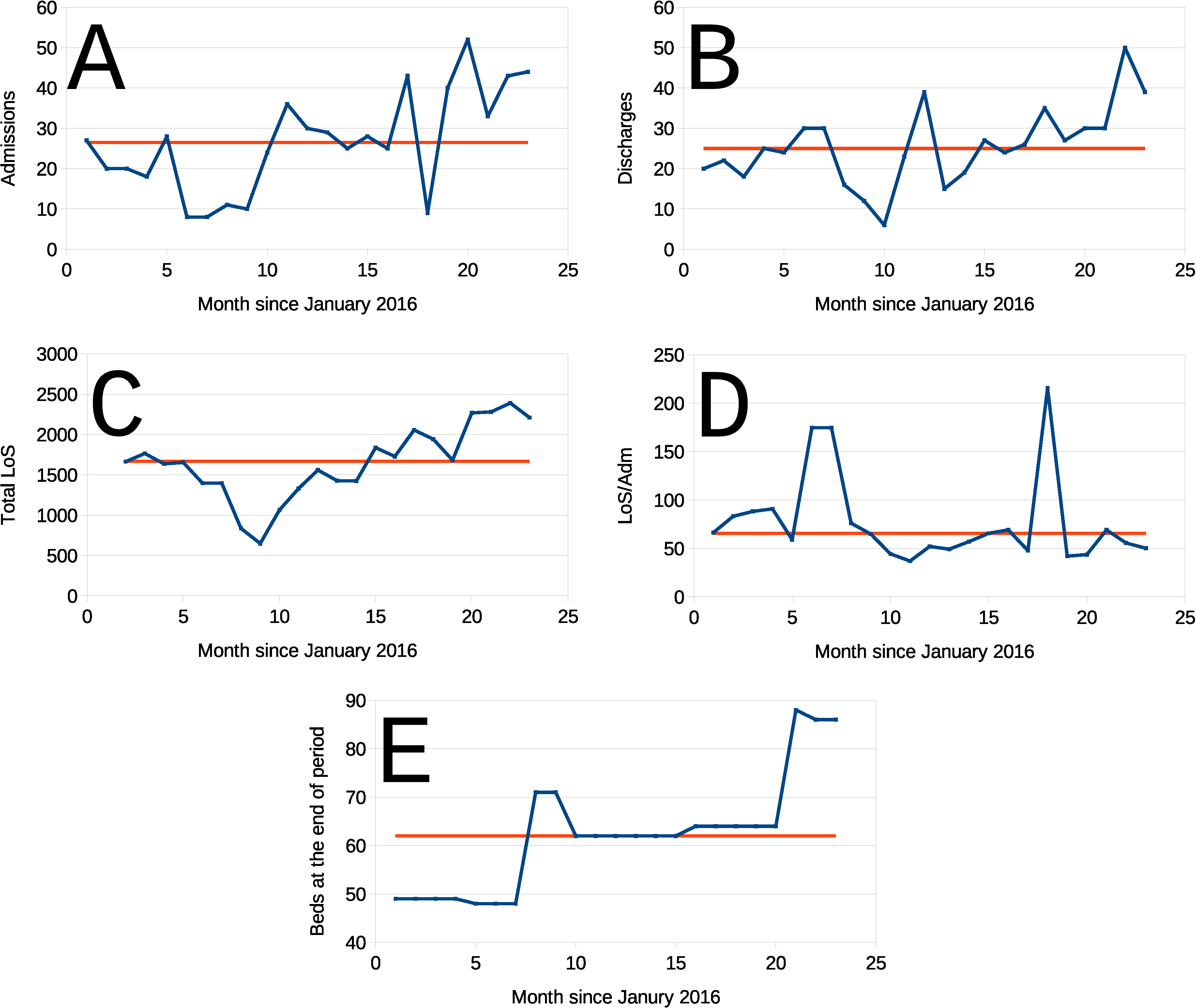}
		\caption{Example of four groups of data from the Function Recovery Uunit. (FRU). \textbf{A}: Total admissions (\textbf{Adm}) sequence ($ n = 22$ , $ n_{runs}  = 11, P_{rrd} \approx 0.15 $); \textbf{B}, Total discharges (Dis) sequence ($ n = 22 $, $ n_{runs} = 8, P_{rrd} \approx 0.04 $); \textbf{C}, is the sum of lengths of stay (\textbf{LoS}) for all in–patients (\textbf{InP}) in the Unit during each month ($ n = 23 $, $ n_{runs} = 5 $, $ P_{rrd} \approx 4.5 \cdot 10^{−4} $); \textbf{D}, is a sequence of the average stay duration (Mean\textbf{ LoS}) of each in–patient at the Unit ($ n = 23 $, $ n_{runs} = 10 $, $ P_{rrd} \approx 0.13$); \textbf{E}, along de observation period the Function Recovery Unit gained relevance and the number of beds were increased, te panel shows the changes in beds at the behind ib several months, as expenses, the analysis of this senescence was found to be non random ($ n = 23 $, $ n_{run} = 5 $, $ P_{rrd} \approx 4.5 \cdot 10^{−4} ) $, actually n runs may be just in this case, the algorithm probably found very minor decimal differences between data an $ \hat{x} $ in one segment; if n $ run s = 3 $, should n runs be $ 3 $ than $ P_{rrd} \l 4.5 ⋅ 10^{−4} $. $ P $ s are probabilities (obtained with the Wald–Wolfovitz test) that the sequences are not randomly distributed about their medians.  As indicated in the text of this communication $ P_{rrd} $ stands for \textit{\textbf{p}robability of \textbf{r}andom \textbf{d}istribution about the median}. In all cases abscissa is time expressed as month number in the sequence, (January 2016 = 1, November 2017 = 23).}\label{F:HospRuns}
	\end{center}
\end{figure}

\subsection{Analysis of runs above and below the median.}

During the period studied, the hospital\textquoteright{s} in–patients were 150 (149 – 151) (median, $ \hat{y} $ , and 95\% CI) ranging (102 – 187) patients. Let us consider a random variable $ y = f (t) $ such that $ f (t) $ is independent from $ f (t − \Delta t) $ or from $ f (t + \Delta t) $ (where $ \Delta t $ is a short time lapse, ideally $ \Delta t \rightarrow 0 $). In other words, a variable for which its present is independent from its past, and whose present does not determines its future. Such variable produces a sequence of data which are uniformly distributed about their median $ \hat{y} $. We say that run occurs when one or more sequence points fall above or below the median, this is easily grasped by looking at Figure \ref{F:HospRuns}. In Figure \ref{F:HospRuns}C there are $ 5 $ runs, and in Figure \ref{F:HospRuns}D are $ 9 $ runs. Points that are equal to $ \hat{y} $ are not taken into account in the analysis. Figure \ref{F:HospRuns} presents four sets of monthly averages from the \textbf{F}unction \textbf{R}ecovery \textbf{U}unit (\textbf{FRU}). Figure \ref{F:HospRuns}A, results from the total of inpatients (\textbf{Adm}) sequence ($ n = 22 $, $ n_{runs} = 11 $, \textit{\textbf{p}robability of \textbf{r}andom run \textbf{d}istribution about the median} $ P_{rrd} $ is $\approx 0.15 $); \ref{F:HospRuns}B, Total discharged patients’ (\textbf{Dis}) sequence ($ n = 22 $, $ n_{runs} = 8, P_{rrd} \approx 0.04 $); \ref{F:HospRuns}C, is the sum of lengths of stay (\textbf{LoS}) for all in–patients (\textbf{InP}) in the Unit during each month ($ n = 23 $, $ n_{runs} = 5 $, $ P_{rrd} \approx 4.5 \cdot 10^{−4} $ ); \ref{F:HospRuns}D, is a sequence of the average stay duration (Mean \textbf{LoS}) of each in–patient at the Unit ($ n = 23 $, $ n_{runs} = 10 $, $ P_{rrd} \approx 0.13 $); \ref{F:HospRuns}E, along de observation period the Function Recovery Unit gained relevance and the number of beds were increased, panel shows the changes in beds at the benign in several months, as expected, the analysis of this sequence was found to be non random ($ n = 23 $, $ n_{runs} = 5 $, $P_{rrd} \approx 4.5\cdot 10^{−4}$ ), actually $ n_ {runs} $ may be just $ 3 $ in this case, the algorithm probably found very minor decimal differences between data \ref{F:HospRuns}an $ \hat{y} $ in one segment; if $ n_{run s} = 3 $, should n runs be 3 then $ P_{rrd} \ll 4.5 \cdot 10^{−4} $. $ P_{rrd}  $s are probabilities (obtained with the Wald–Wolfovitz test) that the sequences are not randomly distributed about their medians. In all cases abscissa is month number in the study period (January 2016 = 1, November 2017 = 23). Out of these $ P_{rrd} $ values the only one indicating a statistically significant non zero number of runs, occurs for \textbf{InP} sequence in \ref{F:HospRuns}C \citep{Benjamin2018, Ioannidis2018}. The non randomness of sequence \ref{F:HospRuns}C could have several reasons, one of them is some kind of \scomillas{periodicity} (a strange one, since its period would be longer that 2 years), but it could also stem from some kind of \textit{random walk}. During the period studied, the hospital\textquoteright{s} in–patients were 150 (149 – 151) (median, $ \hat{y} $ , and 95\% CI) ranging (102 – 187) patients. Let us consider a random variable $ y = f (t) $ such that $ f (t) $ is independent from $ f (t − \Delta t) $ or from $ f (t + \Delta t) $ (where $ \Delta t $ is a short time lapse, ideally $ \Delta t \rightarrow 0 $). In other words, a variable for which its present is independent from its past, and whose present does not determines its future. Such variable produces a sequence of data which are uniformly distributed about their median $ \hat{y} $. We say that run occurs when one or more sequence points fall above or below the median, this is easily grasped by looking at Figure \ref{F:HospRuns}. In Figure \ref{F:HospRuns}C there are $ 5 $ runs, and in Figure \ref{F:HospRuns}D are $ 9 $ runs. Points that are equal to $ \hat{y} $ are not taken into account in the analysis. Figure \ref{F:HospRuns} presents four sets of monthly averages from the \textbf{F}unction \textbf{R}ecovery \textbf{U}unit (\textbf{FRU}). Figure \ref{F:HospRuns}A, results from the total of inpatients (\textbf{Adm}) sequence ($ n = 22 $, $ n_{runs} = 11 $, \textit{\textbf{p}robability of \textbf{r}andom run \textbf{d}istribution about the median} $ P_{rrd} $ is $\approx 0.15 $); \ref{F:HospRuns}B, Total discharged patients’ (\textbf{Dis}) sequence ($ n = 22 $, $ n_{runs} = 8, P_{rrd} \approx 0.04 $); \ref{F:HospRuns}C, is the sum of lengths of stay (\textbf{LoS}) for all in–patients (\textbf{InP}) in the Unit during each month ($ n = 23 $, $ n_{runs} = 5 $, $ P_{rrd} \approx 4.5 \cdot 10^{−4} $ ); \ref{F:HospRuns}D, is a sequence of the average stay duration (Mean \textbf{LoS}) of each in–patient at the Unit ($ n = 23 $, $ n_{runs} = 10 $, $ P_{rrd} \approx 0.13 $); \ref{F:HospRuns}E, along de observation period the Function Recovery Unit gained relevance and the number of beds were increased, panel shows the changes in beds at the benign in several months, as expected, the analysis of this sequence was found to be non random ($ n = 23 $, $ n_{runs} = 5 $, $P_{rrd} \approx 4.5\cdot 10^{−4}$ ), actually $ n_ {runs} $ may be just $ 3 $ in this case, the algorithm probably found very minor decimal differences between data \ref{F:HospRuns}an $ \hat{y} $ in one segment; if $ n_{run s} = 3 $, should n runs be 3 then $ P_{rrd} \ll 4.5 \cdot 10^{−4} $. $ P  $s are probabilities (obtained with the Wald–Wolfovitz test) that the sequences are not randomly distributed about their medians. In all cases abscissa is month number in the study period (January 2016 = 1, November 2017 = 23). Out of these $ P_{rrd} $ values the only one indicating a statistically significant non zero number of runs, occurs for \textbf{InP} sequence in \ref{F:HospRuns}C \citep{Benjamin2018, Ioannidis2018}. The non randomness of sequence \ref{F:HospRuns}C could have several reasons, one of them is some kind of \scomillas{periodicity} (a strange one, since its period would be longer that 2 years), but it could also stem from some kind of \textit{random walk} determined by the uncertainties in $n_j $ and $ \boldsymbol{LoS_i,j} $ \citep{Hastings1993, Sevcik1998a, Sevcik2018}. It is difficult to prove, but the apparent \scomillas{seasonality} was probably determined by the increase in Unit\textquoteright{s} bed capacity depicted in panel \ref{F:HospRuns}E. The figure illustrates that relatively simple sequences of random variables having unknown pdfs such as $ n_j $ and $ \boldsymbol{LoS_{i,j}} $ , may interact in strange manners to produce apparently less random \comillas{simi-periodic} sequences such as figure \ref{F:HospRuns}C.} determined by the uncertainties in $n_j $ and $ \boldsymbol{LoS_i,j} $ \citep{Hastings1993, Sevcik1998a, Sevcik2018}. It is difficult to prove, but the apparent \scomillas{seasonality} was probably determined by the increase in Unit\textquoteright{s} bed capacity depicted in panel \ref{F:HospRuns}E. The figure illustrates that relatively simple sequences of random variables having unknown pdfs such as $ n_j $ and $ \boldsymbol{LoS_{i,j}} $ , may interact in strange manners to produce apparently less random \comillas{simi-periodic} sequences such as figure \ref{F:HospRuns}C.

\subsubsection{La Fuenfr\'{\i}a Hospital data statistical characteristics.}\label{S:FuncEmp}

Figure \ref{F:Admissions_Discharges} presents sequences built in the preceding Subsection \ref{S:Admissions_Discharges}. Panels \ref{F:Admissions_Discharges}A (\textbf{Adm}) and \ref{F:Admissions_Discharges}C (\textbf{Dis}) are sequences with a few negative, artefactual (see Subsection \ref{S:Admissions_Discharges}), outliers. Panels \ref{F:Admissions_Discharges}B and \ref{F:Admissions_Discharges}D are the same data with the scale magnified to see the positive part and better appreciate  data detail when $ y_t \geqslant0 $.

Testing with  the Jarque-Bera \cite{Bera1981a}, the robustified Jarque-Bera-Gel \cite{Gel2008}  and the Shapiro-Wilk \cite{Shapiro1965} tests, we found that none of the data sequences collected  during the study period from La Fuenfr\'{\i}a Holpital, were Gaussian (also called normal) variables, (Probability of Gaussianity, $ P \ll 10^{-6}$). \textbf{InP} sequences contained a large number of ties (points with same value). In ti \textbf{InP} sequence every value is repeated 
\begin{equation*}
	\dfrac{n_{total}}{InP_{max}-InP_{min}} = \dfrac{1329}{187-102} \approxeq 16 \text{ times}.
\end{equation*}

 In the expression: $ \boldsymbol{InP_{max}} $, is \textbf{InP}\textquoteright{}s maximum values; $ \boldsymbol{InP_{min}} $, is \textbf{InP}\textquoteright{}s minimum values and  $ n_{total} $, is total number of points in the sequences. 
Given the narrow range there are only  $ 187-102 = 85$ possible different values among $ 1329 $ integer data in the series, they are not all equally likely. Please  notice that the number of ties increases as $  \boldsymbol{InP_{range}} \rightarrow 0 $, that is the more uniform bed occupation gets, the larger the number of ties, in the \textbf{InP} sequence, will get. This is to be expected when integers are sampled in a narrow range, under these conditions any statistical test, parametric or not,  must be taken with caution, since test power decreases wit ties. Tie corrections were used for the nonparametric statistical procedures employed here \cite{Hollander1973}. Since the Hospital has 192 beds, data indicate that the Hospital operates with some leeway, having an occupancy ranging from $ 53.1 $ to $ 97.3 $\%, with a $ 78.1 $\% median.

With the data in the sequence of \textbf{InP} (shown in Figure \ref{F:Admissions_Discharges}A)  and Eq. (\ref{E:SecDifGen}) a curve of \textbf{DInP} at the Hospital was built, this is shown in Figure \ref{F:DailyHospitalised}B. The sequence in Figure \ref{F:DailyHospitalised}D was obtained by subtracting to each \textbf{Adm} the same day \textbf{Dis}. Figure \ref{F:DailyHospitalised}D presents a sequence obtained by subtracting \textbf{Dis} -  \textbf{Adm} (\textbf{DDiAd}) to the Hospital.

A direct visual comparison between sequences in Figures \ref{F:DailyHospitalised}B (\textbf{DInP}) and \ref{F:DailyHospitalised}D (\textbf{DDiAd}) shows, that in spite of some differences, they are similar, and very different from the series in Figure \ref{F:DailyHospitalised}A (\textbf{InP}). Sequences in Figures \ref{F:DailyHospitalised}B (\textbf{DInP}) and \ref{F:DailyHospitalised}D (\textbf{DDiAd}) compared with the Smirnov test had the same distribution, that is, their distributions are statistically indistinguishable (Probability of identity $ P \lessapprox 1 $, $ n=1329 $, two tails). Yet, when Smirnov test was used to compare sequences \ref{F:Admissions_Discharges}A and  \ref{F:Admissions_Discharges}C, admission (\textbf{Adm}) and discharges (\textbf{Dis}) sequences  were found to be highly statistically significantly different (Probability of identity $ P \ll 10^{-6} $, $ n=1329 $, two tails).

Empirical Hospital occupancy \underline{P}robability \underline{D}istribution \underline{F}unctions (PDF) \cite{Wilks1962} calculated from these data is shown in Figure \ref{F:PDFS}. Daily hospital occupancy data set was found to be not Gaussian when tested with the Jarque-Bera \cite{Bera1981a}, then robustified Jarque-Bera-Gel \cite{Gel2008}  and the Shapiro Wilk \cite{Shapiro1965} tests ($ P \ll 10^{-6} $, in the three cases).  The perception of these data  lack of Gaussianity  becomes intuitive when data  (dots) are observed very dispersed around a Gauss pdf line in the same panel. The Gauss pdf line (shown also as pdf, in Figure \ref{F:DailyHospitalised}C) with the same mean ($ \bar{y} $) and variance [$ s^2(y) $] as the set of 1329 data points forming the sequence in Figure \ref{F:DailyHospitalised}A ($ \bar{y}=149.2 $, $ s[y]=24.4 $).

\subsubsection{La Fuenfr\'{\i}a  Hospital daily admissions and discharge variations sequences.}

\begin{figure}[h!]
	\begin{center}
		\includegraphics[width=12cm]{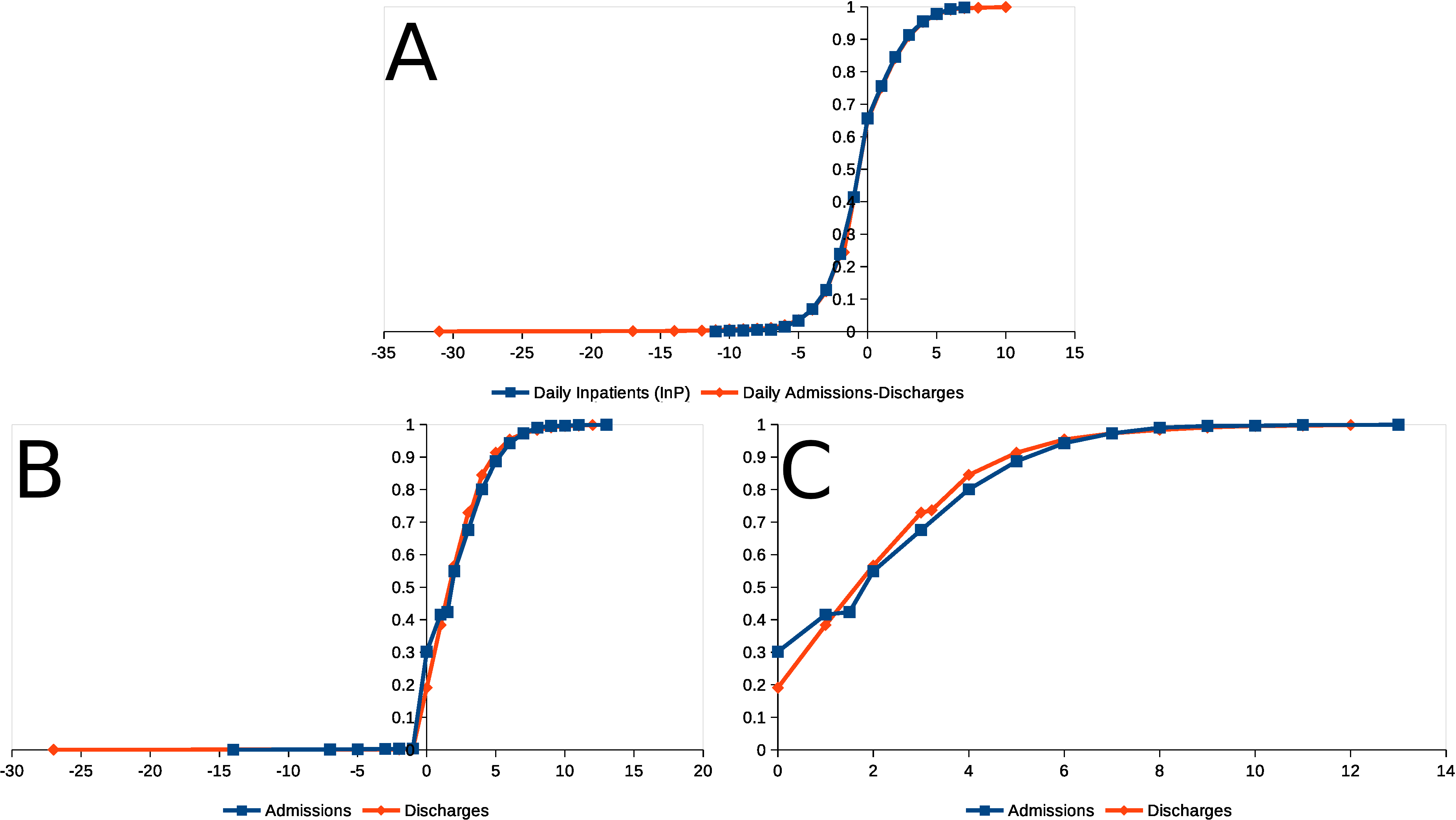}
		\caption{\textbf{Empirical \underline{P}robability \underline{D}istribution \underline{F}unctions  (PDF) \cite{Wilks1962} of daily admissions and stays at La Fuenfr\'{\i}a Hospital from May 12014 to December 19 017.}  \textbf{Panel A}:
		Blue ine with $\blacksquare$ PDF of LFH in-patients (\textbf{InP}  in Figure \ref{F:DailyHospitalised}A); Red line with $\blacklozenge$, PDFs for daily differences in admissions (\textbf{Adm}) and discharges (\textbf{Dis}), called \textbf{DDiAd} in Figure \ref{F:DailyHospitalised}D. \textbf{Panel B}: Blue line with $\blacklozenge$ PDF of LFH daily Admissions in LFH (called \textbf{Adm} in Figure \ref{F:Admissions_Discharges}A); Red line with $\blacksquare$ PDF for LFH daily Discharged \textbf{Dis} in Figure \ref{F:Admissions_Discharges}C). \textbf{Panel C}: Same as Paanel B, but with abscissa expanded to exclude outliers. In all cases abscissa is the value ve aare interested in, and ordinate is the estimated accumulated probability from $ -\infty $ up to the abscissa value.}\label{F:PDFS}
	\end{center}
\end{figure}

Figure \ref{F:PDFS} contains empirical PDF \cite{WikipediaEmpDis2017} calculated for inpatients (\textbf{InP}),  admissions (\textbf{Adn}), discharges (\textbf{Dis}) and difference between \textbf{Adm} and \textbf{Dis}: (\textbf{DAdDis}). It is perhaps good to point out that \textbf{Adm} and \textbf{Dis} were \textit{were logged  daily as separated entries} in Hospital\textquoteright{}s Command Panels. Data in Figure \ref{F:PDFS}  (especially the expansion presented in \ref{F:PDFS}C) indicates that daily \textbf{DInP} is distributed following the \textit{same PDF as \textbf{DAdDis}}, but that \textbf{Adm} and \textbf{Dis} have \textit{different PDFs}. The \textit{largest difference} is observed for probabilities of \textit{no admissions}  (blue line in \ref{F:PDFS}C) and \textit{no discharges} occur (red line in \ref{F:PDFS}C) in a given day (at zero value in the abscissa in Figure \ref{F:PDFS}C), it is  a day without admissions $ 30\% $ likely but a day without discharges is only $ 20 \% $ likely. When the abscissa is $ >0 $, the probability of daily admissions is surpassed by daily discharges. The largest difference between the curve predicted determines the significance of the difference between two PDFs by using the Smirnov test \cite{Smirnov1939, Smirnov1948} (based on Kolmogorov statistics \cite{Kolmogorov1933}). Yet from looking at Figure \ref{F:PDFS}C, it is evident that this is not the only difference between the curves.

\subsection{Daily fluctuations in stay duration and average stay duration in operative units at La Fuenfr\'{\i}a Hospital.}\label{S:DesgloseUnid}

\subsubsection{Random fractals.}\label{S:RandFract}

There are two curves particularly relevant for the analysis of sequences in Figures \ref{F:DailyHospitalised}  and \ref{F:Admissions_Discharges}, they are the, so called, white and Brownian noises \cite{Hastings1993}. In white noise every point in the sequence is independent from de past, and does not determine the future of the system. In brown noise, every point is dependent on the past and determines the future, but in in unpredictable manners.

\paragraph{White noise properties} White noise is a sequence of independent random variables, no value is predictable nor dependent on on any prior or posterior value in the sequence. White noises have a fractal dimension $ D_h=2 $ \cite{Hastings1993, Sevcik1998a, Sevcik2018}. On a first look, a white noise curve looks very similar to sequences in  Figures \ref{F:DailyHospitalised}B and \ref{F:DailyHospitalised}D, the random variable producing each point may have diverse PDFs, white noises are randomly distributed about 0, when $ N \rightarrow \infty$ white noise variables in a sequence produce $\tfrac{N}{2}+1$ runs \cite{Wald1940}. When subject to Fourier transformation \cite{Welch1967} white noises have a constant spectrum for all frequencies between 0 and $\infty$, reminding \textit{white light} which contains all visible electromagnetic frequencies, leading to call this noise \textit{white}. 

\paragraph{Brownian noise properties} Brownian noise is also called \textit{Brown noise} \cite{Brown1828, Brown1829}, \textit{red noise} or \textit{\textit{random walk (Rw)}} \cite{Hastings1993}. For Brownian it  noise always holds that  
\begin{equation}\label{E:BroGen}
f(x)=
\begin{cases}
	0 \text{ if } x=0 \quad \text{(a convention for th observation\textquoteright{}s origin)}\\
	f(x - \Delta x) + \upsilon(\mu=0,\;\sigma^2),
\end{cases}
\end{equation}
where $ \upsilon(\mu=0,\;\sigma^2) $ is a random variable with mean $ \mu =0$ and variance $ \sigma^2 $. Variable $\upsilon$ may be a uniform U[-1,1] pdf, like
\begin{equation}\label{E:Uniform}
U[-1,1]=
\begin{cases}
	0 &\text{ if } x <-1\\
	0.5 &\text{ if } -1 \leqslant x \leqslant 1\\
	0 &\text{ if } x >1.
\end{cases}
\end{equation} 
But very often the variable is a Gaussian (also called \textit{normal}) of the kind $N(0,\;\sigma^2)$ having mean $ \mu = 0 $ and variance $ \sigma^2 $ \cite{Sevcik1998a, Hastings1993}. Brown noise is a particular case of processes called  Markov chains \cite{Rosenblatt1974}, specially important in the so called \textit{waiting time theory} witch studies situations with repressed queues waiting to get some service \cite{BharuchaReid1960, Ross1996}. In layman\textquoteright{}s language, Brownian variables strictly depend on their past, yet they are random, it is impossible to predict their future or estimate their past values. Even though, their current value depends on their past, and their future value depends on the present.

Brown noise in hospital parameters is not unexpected, the admission of a patient depends on bed availability, the amount of beds depends on many criteria of the public health system. Availability also depends on demand, the disease a patient using the bed suffers, the efficiency of doctors in the hospital, treatment availability and so on.  Brownian noise is characterised by wide and slow variations, and has fractal dimension $ D_h=1.5 $ [Eq. (\ref{E:Hausdorff})] and so does $ D_s $ [Eq. (\ref{E:D_h})] when $ N \rightarrow \infty $ \cite{Hastings1993}. Brownian noise distributes around its median (which is not necessarily 0) resulting in a number of runs $\ll \tfrac{N}{2}+1$. 

\subsubsection{Fractal analysis theory of sequences studied here.}\label{S:TeorAnFrac}

As shown in Subsection \ref{S:IntFracAnal}, in contrast with other methods to estimate the fractal dimension  $ D_h $, Eq. (\ref{E:D_h}) is extremely easy to calculate and certainly converges towards $ D_h $ as $ N \rightarrow \infty $, yet though nobody knows how far infinity is for certain \cite{Infinity2020}, we all belie it is far away. Fractal series studied here have an extent which is hard to classify. An $ N=1329  $ points may be long or short depending in somebody\textquoteright{}s point of view, it seems large for conventional statistics, but is short when considered in fractal analysis context \cite{Sevcik2018}. A $ D_s $ value deduced from a 1329 points series underestimates sequence $ D_h $ significantly.  There is a four step solutions for this:
\begin{enumerate}
\item Generate a significant number of sequences with the characteristics of interest using Monte Carlo \cite{Metropolis1987} simulation.
\item Calculate their $ D_s $ and $ \text{var}(D_s) $ [Eqs. (\ref{E:D_h}) and  (\ref{E:varD})] \cite{Sevcik1998a}.
\item Calculate the set of traces\textquoteright{} mean $ D_s $ and its variance including variance between simulated traces as indicated in \cite{Sevcik1998a, DSuze2010a, DSuze2015a, Sevcik2018}.
\item Compare real and simulated sequences parameters using the Vysochanskij--Petunin inequality (details Subsection \ref{S:VysPet} and in \cite{DSuze2010a, Sevcik2018}). 
\end{enumerate}

\subsection{Fractal analysis of sequences from La Fuenfr\'{\i}a Hospital.}\label{S:AnFracIngEg}

\subsubsection{Ruling out sources of arrow in sequences.}
To analyse data in Table \ref{T:FracDimEst} we must take into account the transcription data errors in the SELENE\textsuperscript{\textregistered{} }data base  and how were they corrected for this study, as indicated in Subsection \ref{S:Admissions_Discharges} and the problem of ties mentioned in Sub-Subsection \ref{S:FuncEmp}, also the narrow data rage determines further underestimations of $ D_h $. 

A case random noises in found in \cite{Sevcik2018} where much longer sequences needed for $ D_s $ to converge towards $ D_h $, these sequences contain only decimal digits $ \{0, 1,\ldots,8,9\}  \in \mathbb{Z} $, not real numbers $ \{ -\infty < x < + \infty \} \in \mathbb{R} $.  Empiric data on Tables \ref{T:FracDimSer} and \ref{T:FracDimEst}, may differ from $ D_s  $ of a calibration white noise from a longer sequences. 

\subsubsection{Estimates of the fractal dimension.}

$ D_s $ values estimated for Hospital sequences in Figures \ref{F:DailyHospitalised} and  \ref{F:Admissions_Discharges} are given in Table \ref{T:FracDimSer}. White noise standard set of 100 series with 1329 point s had  an estimated $ D_s=1.5983 \; \pm \; 0.0104$ $\left[ \bar{D}_s \; \pm \; s(\bar{D}_s) \right]$; while  the Brown noise standard set of 100 series with 1329 points had $ D_s=1.2885 \; \pm \; 0.0384$ $\left[ \bar{D}_s \; \pm \; s(\bar{D}_s) \right]$. 

\subsubsection{Statistical compassion of $ D_s $ values appearing in Table \ref{T:FracDimSer}.} 

$ D_s $ values calculated or several white and Brown noises, as well as for La Fuenfr\'{\i}a Hospital parameters discussed previously, are summarized in Table Table \ref{T:FracDimSer}, and $ P $ values to test statistical significance of differences between them appear in Table \ref{T:FracDimEst}.  An asterisk indicates marginally significant $ P $ which could have been more significant if we knew $ D_s $\textquoteright{s} and the Vysochanskij--Petunin inequality would not have been needed to test for significance  \cite{Vysochanskij1980, Vysochanskij1983}, we consider this cases as of undecided significance.  

In--patients sequence $ D_s $ ($ \boldsymbol{ InP}$, Tables \ref{T:FracDimSer} and \ref{T:FracDimEst}, and Fig. \ref{F:DailyHospitalised}A)  is not statistically different from from Brownian ($ P>0.13 $). An asterisk indicates marginally significant$ P $ which could perhaps have been more significant if the Vysochanskij--Petunin inequality use would not have been necessary \cite{Vysochanskij1980, Vysochanskij1983} due to the, we consider this cases as of undecided significance.  

An interesting case is the \textit{Rw} $ \boldsymbol{ \Xi_i} $ (Figure \ref{F:DailyHospitalised}E) built for \textbf{DDiAd} (Figure \ref{F:DailyHospitalised}D) using a recursive  relation
\begin{equation}\label{E:DDiAd}
\boldsymbol{ \Xi_i} = 
\begin{cases} 
	0  &\implies i=1 \\
	DDiAd_i + \boldsymbol{ \Xi_{i--1}}&\implies 2 \leqslant i  \leqslant 1328
\end{cases}
\end{equation} 
Please note the resemblance of sequence $ \Xi_i $ (Fig. \ref{F:DailyHospitalised}D) and sequence in Fig. \ref{F:DailyHospitalised}A . Eyeball inspection of these two series shows tiny differences, yet estimates of their fractal dimension presented in Table \ref{T:FracDimSer} are not identical, $ D_s $ for \textbf{InP} sequences is $ 1.33806 \pm 0.0032$  $\left[ \bar{D}_s \; \pm \; s(\bar{D}_s) \right]$  calculated with Eqs.  (\ref{E:D_h}) and (\ref{E:varD}), respectively, and is $ 1.33922 \pm 0.00382$ for the $ \boldsymbol{\Xi_i} $ sequence ($ N=1329 $ in both cases). 

Fig. \ref{F:DailyHospitalised}A and  Fig. \ref{F:DailyHospitalised}E are not identical, but yet their $ D_s $ values are statistically indistinguishable using Vysochanskij--Petunin [See Equation  (\ref{E:P_VP2}) in Subsection \ref{S:VysPet}]. We have 
\begin{equation}\label{E:ParamDif3A3D}
\begin{split}
	&\Delta D_s =D_{s,InP} -D_{s,\Xi} =1.33922-1.3380 \approxeq 1.16 \cdot 10^{-3}\\
	&s(\Delta D_s) =\sqrt{s^2(D_{s,InP})+s^2(D_{s,\Xi})}
	\approxeq 5.5\cdot10^{-3}\\
	&\then	\lambda=\tfrac{\Delta D_s}{s(\Delta D_s)}\approxeq 0.209662 < \left( \sqrt{\tfrac{8}{3}} = 1.63299\ldots\right) 
\end{split}
\end{equation}
which indicates that Theorem \ref{Te:VysoPet} does not hold to this case, and that Vysochanskij--Petunin inequality, strictly, cannot be used to asses $ P(\Delta D_s) $. Yet by virtue of Corollary \ref{Co:AtLeast} Eq. (\ref{E:AtLeast}) indeed $ P \gg \tfrac{1}{6}\approxeq0.1\bar{6} $ for  $ \Delta (D_s) = 0 $ occurs per chance: i.e., it is \textit{not statistically significant} \cite{Fisher1935c,Benjamin2018, Ioannidis2018}.
\newline

\noindent\wasytherefore \textit{ The \textbf{InP} \textit{Rw} fluctuations were a result of delays between daily Hospital\textquoteright{s} discharges and admissions.} \APLbox
\newline

\begin{table}[h!]
\begin{center}
	\caption{{Statistical comparison between fractal dimensions  ($ D_s $) estimated for the sequences appearing in Table  \ref{T:FracDimSer}}. }\label{T:FracDimEst}
	\begin{tabular}{ccccccc}
		\hline \hline
		& 
		\textbf{Dis}&
		\textbf{Adm}&
		\textbf{DInP} &
		$ \boldsymbol{\Xi_i} $ &
		\textbf{Bro} \textbf{\textsuperscript{\S}}&
		\textbf{Whi}\textbf{\textsuperscript{\S}}\\
		\hline
		\textbf{InP}&
		$ 2\cdot10^{-3} $&
		$ 4\cdot10^{-4} $&
		$ 2\cdot10^{-4} $&
		$\boldsymbol{ \tfrac{1}{6}\leqslant P \leqslant 1}$\textbf{\textsuperscript{\textdied}}&
		$\boldsymbol{0.269}$&
		$7\cdot10^{-4}$\\
		\textbf{Dis}&
		$\dots$&
		$ 2\cdot 10^{-3} $&
		$5\cdot10^{-4}$&
		$ 2\cdot10^{-4} $&
		$ 0.039 $\textbf{\textsuperscript{\ddag}}&
		$2\cdot10^{-3}$\\
		\textbf{Adm} &
		$\dots$&
		$\dots$&
		$ 2\cdot10^{-3} $&
		$ 5\cdot10^{-4} $&
		$ 0.015 $\textbf{\textsuperscript{\ddag}}&
		$ 0.0053 $\textbf{\textsuperscript{\ddag}}\\
		\textbf{DInP}&
		$\dots$&
		$\dots$&
		$\dots$&
		$2\cdot10^{-4}$&
		$9\cdot10^{-3}$\textbf{\textsuperscript{\ddag}}&
		$0.045$\textbf{\textsuperscript{\ddag}}\\
		$ \boldsymbol{\Xi_i} $ &
		$\dots$&
		$\dots$&
		$\dots$&
		$\dots$&
		$ \boldsymbol{0.257}$&
		$8\cdot10^{-4}$\\
		\textbf{Bro} \textbf{\textsuperscript{\S}}&
		$\dots$&
		$\dots$&
		$\dots$&
		$\dots$&
		$\dots$&
		$ 7\cdot10^{-4}$ \\
		\hline \hline
	\end{tabular} 
	\begin{center}
		\textbf{Statistic significances}: calculated using Eq, (\ref{E:VysoPetVP}).  \textbf{Labels mean}: \textbf{InP},in patients at  La Fuenfr\'{\i}a Hospital (Fig. \ref{F:DailyHospitalised}A); \textbf{Adm}, admissions to La Fuenfr\'{\i}a Hospital (Fig. \ref{F:Admissions_Discharges}A); \textbf{Dis}, discharges from La Fuenfr\'{\i}a Hospital (Fig. \ref{F:Admissions_Discharges}D);  \textbf{InP}, daily sequence of inpatients (Fig. \ref{F:DailyHospitalised}A); \textbf{DInP}, InP daily variation ( Fig. \ref{F:DailyHospitalised}B) ; $ \boldsymbol{\Xi_i} $ \textit{Rw} (Fig. \ref{F:DailyHospitalised}E) built using Eq. (\ref{E:DDiAd}); \textbf{Bro}, Brownian noise calculated as indicated in Sub-subsection \ref{S:RandFract} and Eq. (\ref{E:BroGaus}). \textbf{Whi}, White noise calculated as indicated in Sub-subsection \ref{S:RandFract}.   \textbf{\S}:  $ D_s $ is the average of  $ M=100 $ sequences with $N= 1329 $ evaluated with Eq. (\ref{E:D_h}), total variance was  calculated \cite{Sevcik1998a, Sevcik2018}; \textbf{\textbf{{\ddag}}}, are values having borderline significance, but that could more significant if the pdf of data would have been known and the Vysochanskij--Petunin inequality use would had not been necessary, we call them \scomillas{undecided significance}.\textbf{{\textdied}}: Referred to Corollary \ref{Co:AtLeast} for Theorem \ref{Te:VysoPet}.
	\end{center}
\end{center}
\end{table}

\begin{figure}[h!]
	\begin{center}
		\includegraphics[width=12cm]{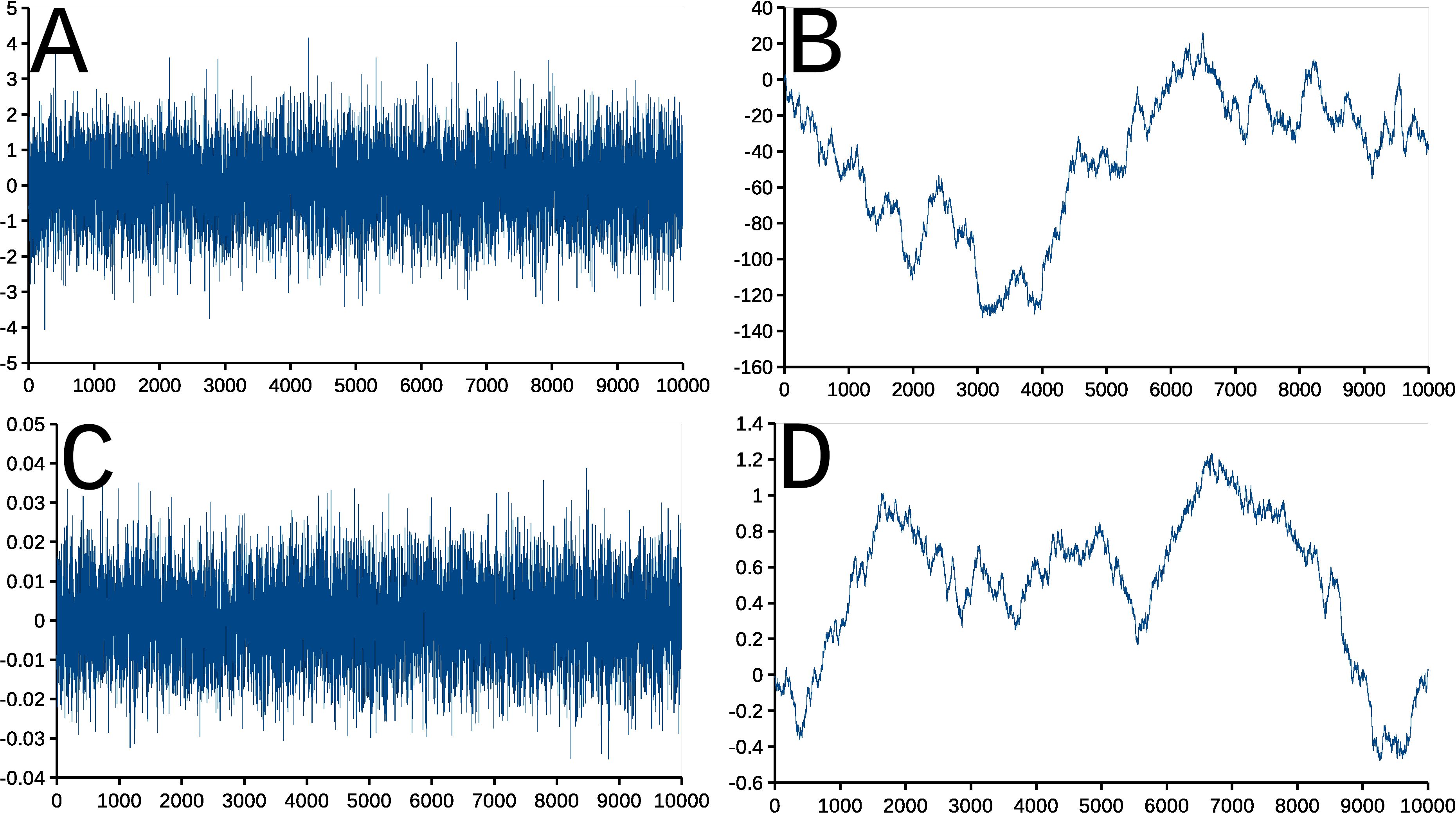}
		\caption{\textbf{Two sequences generated with Monte Carlo simulation using Eq. (\ref{E:BoxMuller}) as described by Box and Muller \cite{Box1958, Press2007}}. Panels are: \textbf{Panel A}, {Gaussian white noise} and ; \textbf{B}: {Gaussian \textit{Rw}}, a \textit{Rw}  generated using Eq. (\ref{E:BroGaus}) from data on pane A; \textbf{Panel C}, {Gaussian white noise} and $ \{g_i = N(0,10^{-4})\}_{i=1,2, \ldots,10^4} $; \textbf{D}: {Gaussian \textit{Rw}}, a \textit{Rw}  generated using Eq. (\ref{E:BroGaus}) from data on pane C  Abscissa and ordinate are arbitrary, Please notice $\left[ D_s \pm s(D_s) \right] $ estimated with Eqs. (\ref{E:D_h}) and (\ref{E:varD}) are: Panel A, $ 1.66122 \pm 7.60191 \cdot 10^{-4} $; Panel B, $ 1.32692 \pm 7.6926\cdot 10^{-4} $; Panel C, $ 1.6692 \pm 7.64176\cdot 10^{-4} $; Panel D, $ 1.31723 \pm 7.72806\cdot 10^{-4} $. \textit{Rw} stands for \textit{random walk}, more details in the communication text.}\label{F:Gauss} 
	\end{center}
\end{figure}

\subsubsection{Monte Carlo simulated sequences.}

White Gaussian noise in Figure \ref{F:Gauss}A  is a sequence of $ \{g_i = N(0,1)\}_{i=1,2, \ldots ,10^4} $ and \ref{F:Gauss}C of $ \{g_i = N(0,10^{-4})\}_{i=1,2, \ldots ,10^4} $. Brownian noise in Figure \ref{F:Gauss}B is the result of 
\begin{equation}\label{E:BroGaus}
b_i = \begin{cases}
	b_1 = 0\\
	b_{i} + g_{i-1} \implies 2 \geqslant i \geqslant 1329
\end{cases}
\end{equation}
where $ b_i $ is the $ i $\textsuperscript{th} point in Figure \ref{F:Gauss}B and $ g_{i-1} $ is the $ (i-1) $\textsuperscript{th} point in Figure \ref{F:Gauss}A. The required random normal variates of type $ N[0,1] $ were generated as indicated in Section \ref{S:Normal}. In contrast with the Gaussian white noise (Figs. \ref{F:Gauss}A and \ref{F:Gauss}C), Brownian noise sequences have slow and fast meanders and relatively long periods with trends to increase and/or to decrease which are just local tendencies in the segment that we happen to observe (Figs. \ref{F:Gauss}B and \ref{F:Gauss}E), they do not represent any periodicity of the system. Segments like the one in Fig \ref{F:Gauss}B for $ 7\cdot10^3 \leqslant x \leqslant 10^4 $ look very much like the \textbf{InP }curve at La Fuenfr\'{\i}a Hospital in Figure \ref{F:DailyHospitalised}A which also produces a false sensation of periodicity with a slight tendency to decrease, but are just chaotic trends, in spite of the fact that the underlying random process (Figs. \ref{F:DailyHospitalised}A, \ref{F:Gauss}A and \ref{F:Gauss}C) have no periodicities at all. A \textit{Rw} expresses a system out of control, just like epidemics \cite{Mandelbrot1983, Hastings1993, Sevcik1998a}, storms  \cite{Burt1988, lorenc1988, Jarraud1989, Lorenz1963, Dunne2013} or earthquakes \cite{Bhattacharya2009}, which if described with statistical parameters based on a (short) periods of time leads to wrong conclusions.

\subsection{Power spectrum  analyses.}

\begin{figure}[h!]
	\begin{center}
		\includegraphics[width=12cm]{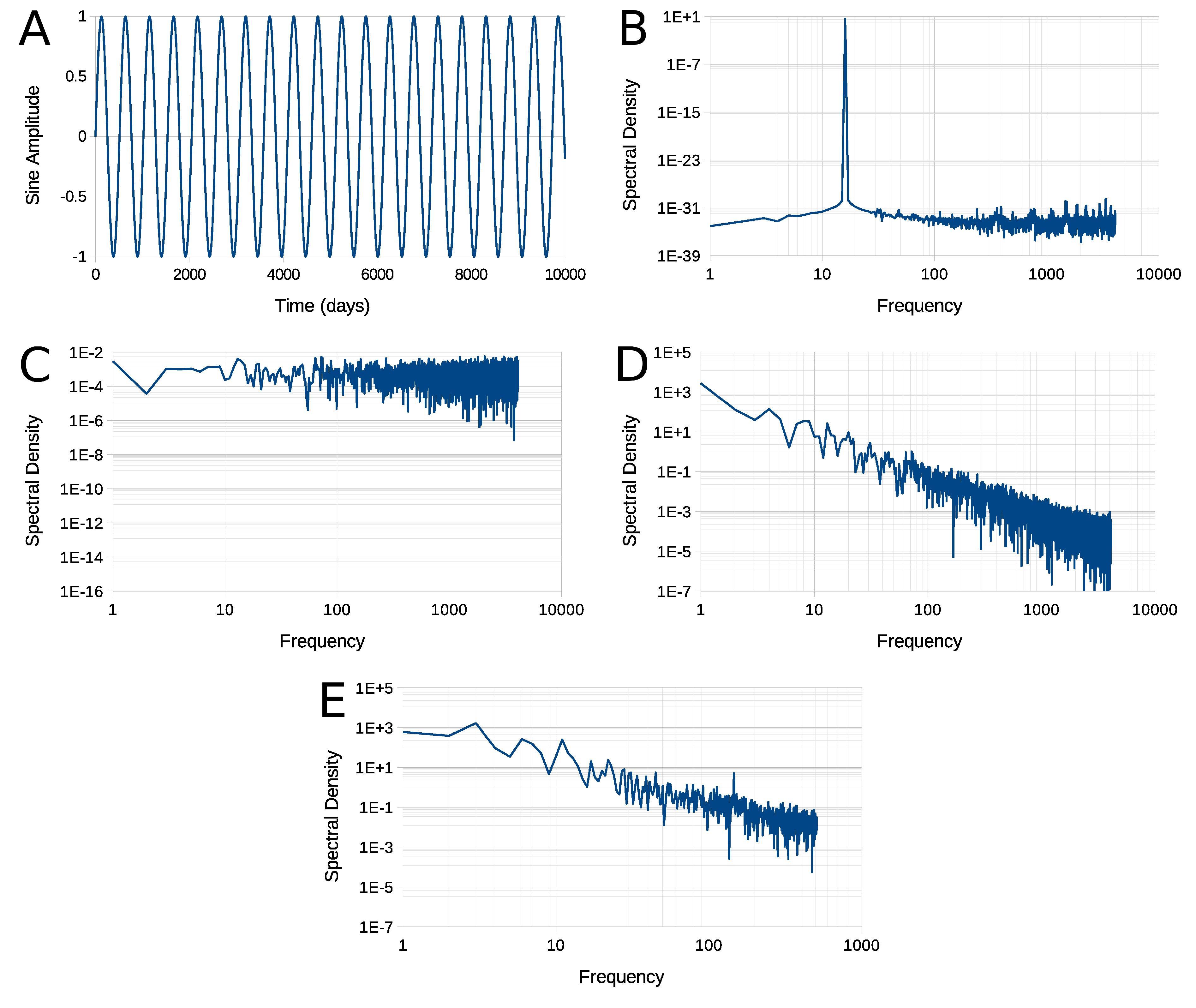}
		\caption{\textbf{False periodicity analysis in \textit{Rw}s. Use of spectral density ($ \boldsymbol{ \psi_{f_j}} $).}  Panels are:  \textbf{Panel A}- A perfectly periodic signal, the sine trigonometric function $ \zeta(x)= \sin(x \tfrac{ \pi }{128}) $; \textbf{Panel B}- $ \psi_{f_j} $ of the sine function shown in panel A; \textbf{Panel C}- $ \psi_{f_j} $ of Gaussian white noise shown at  Figure \ref{F:Gauss}A; \textbf{Panel D}- $ \psi_{f_j} $ of Brown \textit{Rw} shown in  Figure \ref{F:Gauss}B; \textbf{Panel E}- $ \psi_{f_j} $ pf daily inpatients ($ \boldsymbol{ InP} $) sequence which in appears in Figure \ref{F:DailyHospitalised}A. Ordinate of panels  B through E, is power dissipated at each frequency. The sine function depicted dissipates all its energy at a single frequency observed at a single neat peak in its $ \psi_{j} $ (panel B). \textit{The number of hospitalised patients sequence $ SD{f_j} $ corresponds to a Gaussian \textit{Rw} without any underlying periodicity}. All SD plots were filtered using a Hann window \cite{Blackman1959, Harris1978}. The rapid oscillations observed at right of signals in panel B to D is an artefact due to sampling discrete points in the signal for the Fourier transformation. \textit{Rw} stands for \textit{random walk},more details in the  communication\textquoteright{} text.}\label{F:SpectralDensities}
	\end{center}
\end{figure}

A classic form to show function periodicities in time domain, is to transform it them into frequency domain function, which is a sum of sines and cosines, using a fast Fourier transform (FFT) \cite{Blackman1959, Welch1967}. Transformed values for each frequency ($ f $) are \scomillas{complex numbers} ($ \Co $), $ c_f $  comprised of two parts or components: one  is some times referred to just the \scomillas{real component $ a_f $} $ \left(  \Re \right) $ and a, so called, \scomillas{imaginary} $ \left(  \Im \right) $  component ($ b_f  \sqrt{-1}$). A complex number ($ c_f $) is then of the form
\begin{equation}\label{E:CompNumb}
c_f = a_f + b_f \sqrt{-1} = a_f + b_{f}  \boldsymbol{\iota}  \implies
\begin{cases}
	c_f  &\in \{ \Co  \} \\
	a_f  \land b_f &\in \{\Re\} \\
	b_f\boldsymbol{\iota} & \in  \{\Im\},
\end{cases}
\end{equation}
where parentheses indicate sets of numbers, and $ \in $ reads as \comillas{\textit{belongs to}}. The \scomillas{energy}, expressed as either a steady curve\textquoteright{}s non zero curve values, or sharp increases or variations in amplitude (such a as the peak in Figure \ref{F:SpectralDensities}B, something we could informally call its \scomillas{weight}), contributed by each frequency component to total signal energy in a fluctuating complex process is given by 
\begin{equation}\label{E:PowSpect}
\left( \boldsymbol{\psi_f} = \sqrt{a_f^2 + b_f^2}\right)  \in \R
\end{equation}
plotting $ \boldsymbol{\psi_f}  $ against $ f $ id called a \textit{spectral density plot}, and it is areal number.

An intuitive manner to understand this, is to consider a perfect periodic signal such as a trigonometric sine function like $ \zeta(x)= \sin( \tfrac{2 x  \pi}{T}) $ where $ \pi = 3.1415926 \ldots $ and $ T $ is a constant (called  the \textit{waveform\textquoteright{}s period}) with same units (meters, seconds, grams, volts, or else) as $ x $. The example is show in Figure \ref{F:SpectralDensities}A. Figure \ref{F:SpectralDensities}B, presents Figure \ref{F:SpectralDensities}A\textquoteright{}s power spectrum consisting of a single peak $ f = \tfrac{1}{T} $ frequency. Please notice that to calculate $ \boldsymbol{\psi_f} $ the actual signal is sampled in $ \Delta x $ intervals (equivalent to multiplying by something called a \textit{Dirac comb} and the FFT also contains the frequency components of the Dirac comb producing a burst of rapid vibrations in the right half of each trace \cite{Blackman1959}.

Figure \ref{F:SpectralDensities}C is a $ \boldsymbol{\psi_f} $ spectrum of the Gaussian white noise shown in  Figure \ref{F:Gauss}A; ordinates of panels  B through E, in arbitrary units. In Figure \ref{F:SpectralDensities}C it is seen that all frequency have the same \scomillas{weight}, i.e., all frequencies contributions are equal. Figure \ref{F:SpectralDensities}D is a Brown \textit{Rw}  $ \boldsymbol{\psi_f} $ spectrum, ; in double logarithmic coordinates, the spectral density is a straight line which decays with a slope of $ \tfrac{1}{f^2} $. Figure \ref{F:SpectralDensities}E is $ \boldsymbol{\psi_f} $ of daily inpatients ($ \boldsymbol{ InP} $) sequence shown in Figure \ref{F:DailyHospitalised}A.  

Although the  simulated sequences in Figure \ref{F:SpectralDensities} are $ 10000 $ points long and Hospital data are only $ 1329 $ points long, there is an evident similarity between the Brownian noise $ \boldsymbol{\psi_f} $ (Fig. \ref{F:SpectralDensities}D) , and $ \boldsymbol{\psi_f} $ from the sequence of Hospital inpatients ($ \boldsymbol{Inp} $) between May 1 2013 and December 19 2017 (Figure \ref{F:SpectralDensities}E). In both cases $ \boldsymbol{\psi_f} $ are characteristic of a Brown \textit{Rw} and none of these $ \boldsymbol{\psi_f} $s has a peak which could indicate a predominant periodic component which could suggest for any periodicity, in spite of short lapses in the Brown \textit{Rw}s that could give this impression.

Figure \ref{F:SpectralDensities}C is Gaussian white noise $ \psi_{f_j} $ calculated for the signal in Figure \ref{F:Gauss}A. In Figure \ref{F:SpectralDensities}C all frequencies have the same \scomillas{weights}, all frequencies contributions are equal, which leads to the  \scomillas{white} name of this kind of noise.  Figure \ref{F:SpectralDensities}D shows $ \psi_{f_j} $ of the Brownian \textit{Rw} in Figure \ref{F:Gauss}B, it may be seen that in double logarithmic coordinates the spectral density follows a straight line which decays with a slope of $ \tfrac{1}{f^2} $. Finally, also in Figure \ref{F:SpectralDensities}E, we show  $ \psi_{f_j} $ corresponding to daily hospitalised patients sequence shown in Figure \ref{F:DailyHospitalised}A. Except for that simulated sequences in Figure \ref{F:SpectralDensities} are $ 10^4 $ points long and Hospital data are only 1329 points long, there is an evident similarity between the Brownian noise  $ \psi_{f_j} $, and  that $ \psi_{f_j} $ from the sequence of Hospital discharges between May 1 2013 and December 19 2017. In both cases $ \psi_{f_j} $ are characteristic of a \textit{Rw} and \textit{none of these $ \psi_{f_j} $ has no peak which could suggest periodic components which could account for any periodicity, in spite of short periods in the \textit{Rw}s that could give this impression}.

\begin{figure}[h!]
	\begin{center}
		\includegraphics[width=12cm]{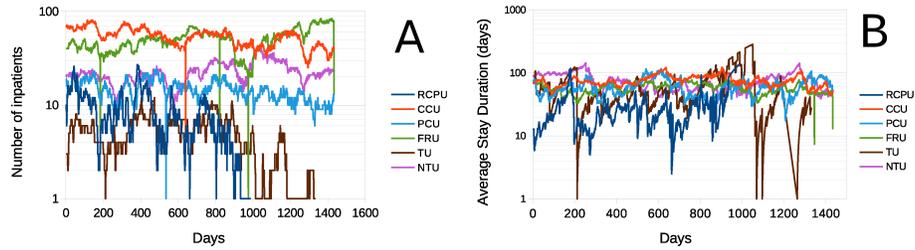}
		\caption{\textbf{Daily fluctuations of in-patients and average duration of stays for each Clinical Operative Unit recalculated every day from January 1 2015 uo to December 12 2017}. \textbf{A}- Number of in–patients (\textbf{InP} ) at each Operative Unit. \textbf{B}- Average stay duration at each Operating Unit [ $ \boldsymbol{\Upsilon_d} $ defined as in Eq. (\ref{E.Upsilon})]. Abscissas are days elapsed from January 1 2014, ordinates are in logarithmic scale to better visualise some of sequence features. Unit initials mean: \textbf{RCPU}, Relapsing Chronic Patients Unit, this unit was suppressed on May 2016 and its beds were transferred to \textbf{CCU}; \textbf{CCU}, Chronic Care Unit; \textbf{PCU}, Palliative Care Unit; \textbf{FRU}, Function Recovery Unit; \textbf{TU}, Tuberculosis Unit; \textbf{NTU}, Neurrehabilitation Treatment Unit.}\label{F:unitFlow}
	\end{center}
\end{figure}

\subsection{Daily fluctuations in stay duration and average stay duration in operative units at La Fuenfr\'{\i}a Hospital.}\label{S:DaylyFluct}

The Hospital is divided into the following Clinical Operative Units (\textbf{COU}): \textbf{RCPU}, Relapsing Chronic Patients Unit, this unit operated until May 2016 when its beds are transferred to \textbf{CCU}; Chronic Care Unit; \textbf{PCU}, Palliative Care Unit; \textbf{FRU}, Function Recovery Unit; \textbf{TU}, Tuberculosis Unit and \textbf{NTU}, Neurorehabilitation Treatment Unit. Figure \ref{F:unitFlow}A presents data on daily number of in–patients (InP ) at each unit, and their average stay duration on Panel \ref{F:unitFlow}B). Daily fluctuations of in–patients and average length of stays (\textbf{LoS}) for each COU calculated every day from January 1 2015 uo to December 12 201. Abscissas are days from January 1 2014, ordinates are in decimal logarithmic scale to better visualise some of sequence large outliers, which are particularly evident in Panel \ref{F:unitFlow}B. Traces in Figure \ref{F:unitFlow} seem some kind of \textit{Rw}. All have some outlying peaks, particularly RCPU and UTB, but, less prominent, occur in other COU too. It must be pointed out that in this paper we some times refer to products or quotients between random variables, this is the case of averages including number of patiens and stay duration in each COU in Figure \ref{F:unitFlow}B. We know that our data are not Gaussian, but even in the case of Gaussian values these averages, may be quotients of random variables following a a “pathological” statistical distribution such as the Cauchy pdf (See Subsection \ref{S:Cauchy}).

\begin{table}
	\begin{center}
		\topcaption{\textbf{Nonparametric linear Theil \citet{Theil1950a, Theil1950b, Theil1950c, Theil1992} regression parameters of LFH clinical units\textquoteright{} time sequences expressed for each unit as in Figure \ref{F:unitFlow}.} }\label{T:FracDimUnid}
		\begin{supertabular}{m{3cm}*{2}{m{4.5cm}}*{3}{m{1.5cm}}}
			\hhline{*{6}=}
			 \textbf{Sequences} &
			\centering $ \boldsymbol{\alpha} $ &
			\centering $ \boldsymbol{\beta} $ &
			\centering $ \boldsymbol{\rho_S} $ &
			\centering $ P $&
			\centering $ N $
			\cr
			\hhline{*{6}-}\\
			\flushright{\textbf{InP}} &&\cr
			\textbf{RCPU} & 
			\centering $ 12.8\; (12.5, 13.0)$ &
			\centering $  -0.010\; (-0.011, -0.009) $ &
			\centering $-0.618$ &
			\centering $ <10^{-6} $ &
			\centering $986$ 
			\cr
			\textbf{CCU} & 
			\centering $ 67.9 \;(67.4, 68.4)$ &
			\centering $ -0.029\; ( -0.021, -0.019) $ &
			\centering $-0.695$ &
			\centering $ <10^{-6} $ &
			\centering $1434$
			\cr
			\textbf{RCPU} & 
			\centering $ 18.3 \;(18.1, 18.4)$ &
			\centering $ -0.005\; (-0.006, -0.005) $ &
			\centering $-0.540$ &
			\centering $ <10^{-6} $ &
			\centering $1434$
			\cr
			\textbf{RFU} & 
			\centering $ 37.7 \;(37.1, 38.2)$ &
			\centering $ 0.019 \; (0.018, 0.021) $ &
			\centering $0.558$ &
			\centering $ <10^{-6} $ &
			\centering $1434$
			\cr  
			\textbf{TB} & 
			\centering $ 6.6 \;(6.5, 6.8)$ &
			\centering $ -0.003 \; (-0.0033, -0.0026) $ &
			\centering $-0.447$ &
			\centering $ <10^{-6} $ &
			\centering $1331$
			\cr
			\textbf{TNU} & 
			\centering $ 18.8\;( 18.5, 19.1)$ &
			\centering $ 0.005 \; (0.004, 0.006) $ &
			\centering $0.423141$ &
			\centering $ <10^{-6} $ &
			\centering $1434$
			\cr
			\flushright{$ \boldsymbol{\overline{\Upsilon_d}} $} &&&\cr
			\textbf{RCPU} & 
			\centering $ 16.1 \; (15.2, 16.9)$ &
			\centering $ 0.018 \; (0.015, 0.021) $ &
			\centering $ 0.343$ &
			\centering $ <10^{-6} $ &
			\centering $985$ 
			\cr
			\textbf{CCU} & 
			\centering $ 67.9 \;(67.4, 68.4)$ &
			\centering $ 0.001\; (-0.0009, 0.0030) $ &
			\centering $0.002$ &
			\centering $ 0.96 $ &
			\centering $1434$
			\cr
			\textbf{CPU} & 
			\centering $ 54.4\;(53.5, 55.4)$ &
			\centering $ 0.006\; (0.004, 0.009) $ &
			\centering $0.123$ &
			\centering $2\cdot 10^{-6} $ &
			\centering $1434$
			\cr
			\textbf{RFU} & 
			\centering $ 59.6\;(59.0, 60.2)$ &
			\centering $ -0.003\; (-0.005, -0.002) $ &
			\centering $-0.107$ &
			\centering $ 4 \cdot 10^{-5} $ &
			\centering $1434$
			\cr  
			\textbf{TB} & 
			\centering $ 53.8 \;(51.6, 56.1)$ &
			\centering $ -0.009\; (-0.013, -0.004) $ &
			\centering $-0.181$ &
			\centering $ <10^{-6} $ &
			\centering $1331$
			\cr
			\textbf{TNU} & 
			\centering $ 84.1 \;(83.0, 85.3)$ &
			\centering $ -0.017 \; (-0.020, -0.014) $ &
			\centering $-0.312$ &
			\centering $ <10^{-6} $ &
			\centering $1434$
			\cr
			\hhline{*{6}=} 
		\end{supertabular} 
	\end{center}
	\begin{footnotesize}
		 Sequence of \textbf{InP}: number of in-patients in the Unit. $ \boldsymbol{\overline{\Upsilon_d}} $ sequences are: average stay duration at each Operating Unit [defined as in Figure \ref{F:unitFlow}  by Eq. (\ref{E.Upsilon})].  Data presented as medians and their 95\% Confidence Interval (between parentheses), $ \boldsymbol{\alpha} $ regression line value when $ x=0 $ (intercept), $ \boldsymbol{\beta} $ straight line regression coefficient (slope), $ \boldsymbol{\rho_S}$ nonparametric Spearman \cite{Spearman1904} crenellation coefficient, $ P $ is the probability that each sequence comes from a population with es $ \rho_S = 0 $, usually called the probability that the correlation is not statistically significant, and $ N $ is the number of data pairs used for the linear regression analysis. HLF unit initials mean: \textbf{RCPU}, Relapsing Chronic Patients Unit, this unit was suppressed on May 2016 and its beds were transferred to \textbf{CCU}; \textbf{CCU}, Chronic Care Unit; \textbf{PCU}, Palliative Care Unit; \textbf{FRU}, Function Recovery Unit; \textbf{TU}, Tuberculosis Unit; \textbf{NTU}, Neurorehabilitation Treatment Unit.
	\end{footnotesize}
\end{table}

\subsection{Operative units nonparametric descriptive statistical analysis of sequences in Figure \ref{F:unitFlow}}

\begin{tiny}
	\begin{tabular}{*{2}{m{8.4cm}}}
		\begin{flushleft}
			\textquotedblleft{}If a sample is too small you can\textquoteright{t} show that anything is statistically significant, whereas if sample is too big, everything is statistically significant.\textquotedblright $ \ldots $  \textquotedblleft{} you are doomed to significance.\textquotedblright{ }\\
			\citet[pg. 24]{Norman1989}.
		\end{flushleft}
		\\
		&
	\end{tabular}
\end{tiny}

As said in Section \ref{S:DaylyFluct} sequences in Figure \ref{F:unitFlow}B include data such \textbf{InP} average patients\textquoteright{} stay duration in each operative unit ($ \boldsymbol{\Upsilon} $). This number calculated as:
\begin{equation}\label{E.Upsilon}
\overline{\Upsilon} =\frac{\sum_{j=1}^{N_d}\Upsilon_{d,j}}{N_d}
\end{equation}
where $ d $ represent a given period and $ j $ is the $ j^t\text{h} $ patient in the service, and $ N_d $ is the number of patients in each service at day $ d $. Thus, $ \Upsilon_d $ are quotients unknown distribution random variables. As mentioned in Section \ref{S:Cauchy} those quotients of random variables have uncertain distributions, which in case of Gaussian variables is the Cauchy pdf Eq. (\ref{E:Cauchy_pdf}) discuses in Section \ref{S:Cauchy}. These quotients follow unknown pdfs when terms in the ratio obey unknown pdfs. For these distributions $ \bar{x} $ n and $ s^2 (x) $ are meaningless. Statistical sampling theory is base on the fact that sample mean ($ \bar{x} $) and variance [$ s^2 (x) $] converge towards population values such a $ \mu $, the population’s mean, and $ \sigma^2 $ , the population\textquoteright{s} variance. But $ \mu $ and $ \sigma^2 $  do not exist for Cauchy and other, so called, \comillas{pathological distributions} for which $ \bar{x} $ and $ s^2 (x) $ have nowhere towards to converge to. Cauchy sample experimental estimates of $ \bar{x} $ and $ s^2 (x) $ vary wildly, specially when sample sizes grow and large outliers outliers values become very likely \citep{Rothenberg1964, Fama1968}. Due to these only nonparametric descriptive analysis provides meaningful information on sample localization and dispersion. A caveat is due here, data samples from La Fuenfr\'{\i}a Hospital, except for data in Figure \ref{F:HospRuns}, are rather large ($ 986 \leqslant N \leqslant 1434 $). Under these circumstances small, but statistically significant changes, which lack relevance or meaning in practice, may be detected, \citep{Theil1992, Lohninger2012, Hojat2004, Runkel2012}. Table \ref{T:FracDimUnid} contains results where sequences in Figure \ref{F:HospRuns} are subjects o a nonparametric Theil regression analysis \citep{Streiner2012}. As seen in Table \ref{T:FracDimUnid}, many of the hospital units sequences in Figure 6 seem to produce a small, but highly statistically significant, correlation between the sequence under study and time. All time series in this figures are \textit{Rw}s (not shown), as was confirmed by their spectral density, characteristic of Brownian noise, without any evidence of periodicity, and resembling those in Figures \ref{F:unitFlow}D and \ref{F:unitFlow}E. As said in connection with the sequence in Figure \ref{F:Gauss}B and \ref{F:Gauss}D, \textit{Rw}s may exhibit, some times rather long, trends to increase or to decrease, that are independent from any variation in factors determining the \textit{Rw}. The pdf of $ \Delta y t = y_t - y_{t−\Delta t} $ does not change in those periods,. Due to these meandering of time sequence, descriptive statistics parameters, even if optimally chosen, like in Table \ref{T:FracDimUnid} fail to provide meaningful information on \textit{Rw}s like Figure 6B and some of the preceding figures.

\section{Discussion.}\label{S:Caveats}

\subsection{The study addresses an empirical finding.}

At least during the study period, a Fuenfr\'{i}a Hospital operates, in median,  at 76.8\% of its total capacity an fluctuates in what could be taken as slow periodic meanders (Figure \ref{F:DailyHospitalised}A). Yet, the daily admissions to (Figure \ref{F:DailyHospitalised}B) or discharges from  LFH (Figure \ref{F:DailyHospitalised}D) are random sequences which show no periodicities whatsoever. Daily differences (Figure \ref{F:DailyHospitalised}D)  between patients admitted (Figure \ref{F:Admissions_Discharges}B) and discharged  (Figure \ref{F:Admissions_Discharges}D) from the Hospital were also random aperiodic sequences. It was somewhat surprising to find that the operating Hospital\textquoteright{s} curve (Fig. \ref{F:DailyHospitalised}A)) could almost perfectly (Fig.  \ref{F:DailyHospitalised}E) reproduced when daily differences between admitted and discharged patients (Fig. \ref{F:DailyHospitalised}D) were added sequentially as indicated in Section\textquoteright{s} \ref{S:RandFract} Eq. (\ref{E:BroGen}).

\subsection{Meaning of a process\textquoteright { }\textit{lack of control}.}\label{S:LackOfControl}

Data on Figures  \ref{F:DailyHospitalised}A  and \ref{F:SpectralDensities}E reflect that Hospital bed occupancy is described by a random fractal function, the so called Brownian \textit{Rw}. Just like Brownian \textit{Rw}, bed occupancy fluctuates with important periods of low occupancy. 

The main characteristic of Brownian \textit{Rw}. is that its complexity (reflected in its $ D_s $) does not change if daily inpatients  (\textbf{DInP}) increases or decreases, this is called self affinity of the process \cite{Mandelbrot1983, Hastings1993}, it does not change with scaling \textbf{DInP}. This is shown by the Monte Carlo simulated traces of Figure \ref{F:Gauss} where the Gaussian white noise $  \{g_i = N(0,1)\}_{i=1,2, \ldots,10^4} $ in panel \ref{F:Gauss}A looks very similar to the $  \{g_i = N(0,10^-4)\}_{i=1,2, \ldots,10^4} $ white noise in panel \ref{F:Gauss}C, exp of course for the ordinate scale; Something similar happens with the associated Brownian noises in panels \ref{F:Gauss}B and \ref{F:Gauss}D, which have similar complexities, but the fluctuations in Panel \ref{F:Gauss}B are approximately 10 times wider in than in Panel \ref{F:Gauss}D. 

\textit{Still, from a production efficiency view point if the differences ( in Figure \ref{F:Gauss}) are very important since the fluctuation range of the Brownian \textit{Rw} diminishes when $ g_i \rightarrow 0$ and production of goods or services increases as the system becomes more predictable (controlled), uniformly productive and efficient.} 

In La Fuenfr\'{\i}a Hospital \textbf{InP} would fluctuate less if the lag between \textbf{Adm} and \textbf{Dis } $\rightarrow 0$. To identify a problem is to start solving it. This means, making the Hospital more efficient requires to analyse the factors that keep \textbf{Adm} and \textbf{Dis} out of synchrony.

\subsection{Hospital La Fuenfr\'{\i}a Characteristics \cite{Rodriguez-Hernandez2017}.}

Hospital La Fuenfr\'{\i}a (\url{https://bit.ly/32FVmyf}) tougher with Virgen de la Poveda Hospital and  Guadarrama Hospital is one of three medium and long stay SERMAS hospitals in Madrid\textquoteright{s} Autonomous Region (MAC)  Sanitary Chancellery (\textit{Consejería de Sanidad}), it is localised in Cercedilla Municipality at the north of Madrid Region, at 60 km from Madrid city. The  MAC comprises an area of of $ 8021.8 $ km$ ^2 $ and a population of $ 6.6  $ M.

The long durations of stay are related to local norms and procedures to handle patients, it is likely that in Spain the public health system has more leeway to tolerate longer stays than perhaps other nations in the EU, specially when dealing with needy patients. Yet, LFH is a place to deal with chronic cases, and it most important unit is FRU, responsible for recovering patients with disabilities stemming from bone surgery (prothesis, fractures, etc.) and motor deficiencies following cerebrovascular accidents (CVA or ictuses). Thus perhaps it is to be expected that stays at LFHa may be longer than in hospitals dealing under ambulatory treatment.

LFH is located in a pleasant rural area, Madrid\textquoteright{s} Sierra de Guadarrama, convenient for its goals, but distant from Madrid (population 3.3 M) city and MAC, in general. Patients are referred from other SERMAS hospitals, thus, delays between LFH discharges and new patients arrivals,  are not surprising. The analyst of data in this paper (CS) was completely unaware of this factors, but they popped out sharply from data analysis discussed here.

\subsection{Classical statistical analysis of institutional efficiency.}

Descriptive statistics, usually using parametric (Gaussian) methods are the most common manner employed tools to gauge institutional efficiency. LFH SELENE\textsuperscript{{\textcopyright}'} include command panels parameters that are ratios of random variables. Those ratios follow unknown statistical distributions which like the Cauchy pdf, have no central moments, mean or variances do not exist, thus textbook recipes like
\begin{equation*}
		\bar{x} = \dfrac{\sum_{i=1}^{N} x_i}{N}  \qquad \text{and/or} \qquad
		s^2(x) = \dfrac{\sum_{i=1}^{N}\left( x_i - \bar{x}\right)^2 }{N-1},
\end{equation*}
for sample mean and variance, are, calculable but, absolutely meaningless \cite{Cramer1991, Pitman1993, Walck1996, Wolfram2003}(See Section \ref{S:Cauchy}). For these kind of random samples the use of nonparametric statistical estimators such as medians and 95\% CI must always be preferred to $ \bar{x} $ and $ s^2(\bar{x}) $ which, are also, very sensitive to outliers and, as said,  are meaningless for Cauychyan variables. Furthermore, in the analysis of data extracted from LFH command panels, we found no Gaussaian variables (Sub-Subsection \ref{S:FuncEmp}), further demanding the use of nonparametric statistical methods.

Another example of statistics failure to productively help to analyse aspects of LFH operative efficiency may be derived from Figure \ref{F:unitFlow}. The figure presents unit operation data are \textit{Rw}s with large outliers. But when these \textit{Rw} were subjects of linear nonparametric regressions (Table \ref{T:FracDimUnid}) and apparent correlations with time appeared, with correlation coefficients statistically distinct from zero, i.e. apparently \textit{highly} unlikely to differ from zero per chance. Large samples  are are good way to discover statistically significant, but irrelevant, trends, which , as shown here, may just temporary tendencies of a random walk. Such correlations could be used to create institutional (probably wrong) operation policies. Classical statistics is notoriously limited to study fractal processes \cite{Mandelbrot1967, Pincus1991, Mandelbrot1983, Hastings1993, DSuze2010a, DSuze2015a}. So, of our  findings regarding statistical evaluation of LFH operational curves is neither new nor really surprising. The novelty of our findings relates to the commonality of statistical analysis and the lack of other analyses (fractal or else) to study institutional efficiency.

Several couple of caveats are needed here. 
\begin{enumerate}
	\item This communication is centred on maximising the use of hospital facilities as an indication of efficiency. Yet, it does not take into account other criteria such as patient\textquoteright{}s welfare, degree of recovery , speed of recovery, which are of fundamental importance, but also depend on efficient hospital resources use.
	\item The many theoretical tool we use is fractal analysis, but to accurately estimate fractal dimension very long data series \cite{Sevcik1998a,  Hasselman2013, Sevcik2018}. this limitation, as done here, may be overcome py the use of sequences having known fractal dimensions as done here an in \cite{Sevcik1998a, Sevcik2018}.
	\item Spectral analysis confirmed the lack of periodicity in the \textbf{InP} time series. 
	\item Yet, the most important finding reported here is that that the wide fluctuations observed in the inpatients (\textbf{InP}, Figure \ref{F:DailyHospitalised}A) may be reconstructed with identical naked-eye and fractal properties ($ \boldsymbol{\Xi_i} $ sequence in Figure \ref{F:DailyHospitalised}E) by adding the daily differences in LFH patients admitted and discharged (\textbf{DDiA}d, Figure \ref{F:DailyHospitalised}D) using Eq. (\ref{E:DDiAd}). \textit{These are empirical finding no possible to result from any mathematical theory or artefact}.
\end{enumerate}
\subsection{\textit{Rw} in La Fuenfr\'{\i}a Hospital operation curves.}\label{S:MarchasAle}

The most surprising finding in this study, is that curves representing in-patients, \textbf{InP}, and \textbf{$ \boldsymbol{\Delta}  $InP} are random Markovian processes \cite{Rosenblatt1974}. In a Markovian processes past determines what happens now, and what occurs now decides the future. If there were no random components in the sequence, known the present would allow to know past and future of the system. But since uncertainty separates an instant from the next, the whole system becomes unpredictable. It dt drifts over time in ways that a naive observer of a short span, is driven to believe that the system is predictable and determined. 

If a present drift trend pleases us, we tend to believe that what ever we are doing is \comillas{correct}, in the opposite case we would tend to believe that we are doing something \comillas{wrong}, even if in reality we have no \comillas{merit}  in the first case nor are we \comillas{guilty} in the second case. Examples of this are seen at times  from 0 to 6000 in Figure \ref{F:Gauss}B. But after that time the \textit{Rw} in Figure \ref{F:Gauss}B which produces a false appearance of relative \comillas{stability}, even when actually at all those times is exactly the same system under the same $ \{g_i = N(0,1)\}_{i=1,2, \ldots,10^4} $ Gaussian conditions as shown by the plot in Fig. \ref{F:Gauss}A.

\section{Caveats and recommendations.}

\subsection{The contribution of our work.}
We know (both authors are trained MDs) that hospitals and medicine are complex, yet our study shows that the waveform meandering between full and insufficiency occupancy at LFH may be perfectly, exactly, explained considering a single factor: admission and discharges are systematically and randomly out of phase. These random difference, small as it may be, is the de terminating factor at LFH: To the best of our knowledge nobody has previously singled this small but important hidden factor.
We know (both authors are trained Mds) that hospitals and medicine are complex, yet our study shows that the waveform meandering between full and insufficiency occupancy at LFH may be perfectly, exactly, explained considering \textit{a single factor: admission and discharges are systematically and randomly out of phase}. These random difference, small as it may be, is the de terminating factor at LFH: To the best of our knowledge nobody has previously singled this small but important hidden factor.

\subsection{Detecting problem indicators.}

Descriptive statistics and common \textit{Command panels}, are not enough to detect hidden administration problems with impact on hospital efficiency. Descriptive statistics limitations mentioned in section Subsection \ref{S:MarchasAle} open a question on how to evaluate and optimise La Fuenfr\'{\i}a Hospital operation and administration whose operation curves are Markovian random walks.

Random walk curves meanders reduce system efficiency. In a hospital performance case, low occupation inefficient periods.
\begin{enumerate}
	\item The population demanding service do not gets it. 
	\item The cost of operating the hospital (infrastructure, salaries, etc.) are wasted at low occupancy periods. 
\end{enumerate}
At la Fuenfr\'{\i}a Hospital low occupation got as low as 45\% (Figure \ref{F:DailyHospitalised}A) of its capacity. As discussed in Section \ref{S:FuncEmp}  oscillations between full and partial occupation seem related to a daily lag between Admissions and Discharges. The analysis indicates that should this lag be eliminated fluctuations in occupancy would disappear.

During the period of this study the mean number of LFH inpatients was $ \boldsymbol{\overline{InP}} = 149.2 $ almost identical to its median $ \boldsymbol{\widehat{InP}} = 150.0 $, and ranged $ (102, 187) $. Since the hospital had 192 patient\textquoteright{s} beds, thus the bed occupancy during the study period was $ \bar{x} = 77.7\% $ and a median of $ \hat{x}= 78.2 $ ranging $ (53.1, 97.4)\% $. Thus over the observation period, on average $ 22\% $ ranging $ (3, 47)\%  $ La Fuendr\'{\i}a Hospital beds were vacant, which means that a similar proportion of the recurses assigned to LFH was lost, and this thus not generates in lack of demand. 

\subsection{Is it possible to solve the problem?}

Our study indicates that LFH vacancies could be eliminated or greatly reduced if the daily difference $ \boldsymbol{Adm} -\boldsymbol{Dis} $ is annulled (Subsection \ref{S:LackOfControl}, Figs. \ref{F:Admissions_Discharges} and \ref{F:Gauss}). Efficient service by an institution characterised by a \textit{Rw} in time may be increased by reducing broad meanders of the walk (Fig. \ref{F:Gauss}). In case of La Fuenfr\'{\i}a Hospital this study suggests that the key factor reducing efficiency is the the daily difference between admissions and discharges. An improvement in synchrony between people in charge of both processes could help a lot. 

Yet, the solution probably transcends, La Fuenfr\'{\i}a Hospital realms. We have no information on other SERMAS hospitals, but the description of LFH made at the Section  \ref{S:Preamble} illustrates, LFH is interwoven in the SERMAS network, thus it seems necessary to consider, and modify communications within the network. The essay part may be to improve communications between admission and discharge department within LFH, although this may imply some changes to the internal computer network making, if possible, that the admissions department to get notice ahead of the actual discharges in order to transmit to the SERMAS network that LFH will have space for a new patient to have it ready to occupy the space immediately when it actually becomes available.

But this immediate availability is also hindered by the large area SERMAS covers, the fact that there is a 60 km distance from Madrid city (the largest set of SERMAS patients), and even more from other areas in Madrid\textquoteright{s} Autonomous Region. Factors such as means and speed of transportation may be critical, institutional transportation within SERMAS network is probably faster and easier to program, than to solely depend on the patient\textquoteright{s} and its family\textquoteright{s} resources and efficiency, for example. But if the efficiency of SERMAS hospitals is affected to the same extent as La Fuenfr\'{\i}a Hospital\textquoteright{s}, it may be well worthwhile to study the system and to implement solutions between admission and discharges in SERMAS hospitals to improve the system\textquoteright{s} efficiency.

\section{Appendices}

\begin{appendices}
\section{Mathematical appendix}

\subsection{Calculating $ D $, approximated fractal dimension of sequences.}\label{S:IntFracAnal}

Classical (Euclidean) geometry is useful to describe relatively simple forms such as straight lines, squares, circles, cones, pyramids, etc. All these forms may be represented as a linear combination of an integer number of orthogonal components called Euclidean dimensions. One- or two-dimensional Euclidean spaces are subsets of our sensory tri-dimensional space. Yet, almost all natural forms cannot be represented in terms of Euclidean geometry.

According to Mandelbrot \cite[pg. 15 and Ch. 39]{Mandelbrot1983}: 
\begin{quote}
	\comillas{A fractal is by definition a set for which the Hausdorff– Besicovitch dimension  strictly exceeds the topological dimension. Every set with a non-integer $ D $ is a fractal.}
\end{quote}

The Hausdorff–Besicovitch dimension of a set in a metric space may be expressed as \cite{Mandelbrot1983}:
\begin{equation}\label{E:Hausdorff}
	D_{h}=-\underset{\epsilon \rightarrow 0}{\lim }{\tfrac{\ln [N(\epsilon )]}{\ln (\epsilon )}} = -\underset{\epsilon \rightarrow 0}{\lim } \log_{\epsilon}[N(\epsilon)]
\end{equation}
where \textit{N(${\epsilon}$)} is the number of open balls or radius \textit{${\epsilon}$} required to cover the set. In a metric space given any point $ X $, an open ball centred on $ X $ with radius ${\epsilon}$, is the set of all points $ x $ for which distance between $ X $ and $ x $ is $ <\epsilon $.

The term waveform is applied to the shape of a set, usually drawn as instant values of a periodic quantity in time. Besides classical methods such as moment statistics and regression analysis, properties like Kolmogorov y Sinai entropy \cite{Grassberger1983}, apparent entropy \cite{Pincus1991, Pincus1991a, Pincus2001} and fractal \cite{Sevcik1998a} have been proposed to analyse waveforms. Fractal analysis may provide information on spacial extent (tortuosity or capacity to extend in space) and on self-similarity  (the property of staying unchanged when measure scale changes ) and self affinity \cite{Barnsley1993}. In bidimensional spaces waveforms are planar curves, which may have coordinates of different units or dimensions.

There is a simple algorithm to approximate fractal dimension of a curve \cite{Sevcik1998a}. To achieve this, fractal dimension $ D_s $ is estimated for a set of $N$  points from a set$\{x_i,y_i\}_{i=0,1,2,\ldots,N}$ values were the abscissa $x_i$ increases by a constant factor ($\Delta x$).  The curve is transformed into a unit square (a square with side length equal to 1) \cite{Sevcik1998a} as follows. A first transformation normalises every abscissa point as:
\begin{equation}\label{S:x_transf}
	x_{i}^{\text{*}}=\frac{x_{i}}{x_{\mathit{max}}}
\end{equation}
where $ x_i $s are the original abscissa values,  $ x_{max} $ is the largest ${x_i}$. A second transform normalises the ordinate as:
\begin{equation}\label{S:y_transf}
	y_{i}^{\text{*}}=\frac{y_{i}-y_{\mathit{min}}}{y_{\mathit{max}}-y_{\mathit{min}}}
\end{equation}
where $y_{i}$ are the original ordinate values, and $y_{min}$ and $y_{max}$ are minimum and maximum $y_{i}$, respectively. The unit square may be seen as covered with a ${N{\cdot}N}$ grid of cells. $ N $ containing a transformed curve point. A length $ L $ line may be divided into $N(\epsilon)=L/(2 \cdot \epsilon)$ segments of mean length $2 \cdot\epsilon$, and covered with $ N $  radius $\epsilon$ open balls. Fractal dimension is the approximated as $ D_s $ \cite[Ecuation (6a)]{Sevcik1998a} as
\begin{equation}\label{E:D_h}
	D_{h} = \underset{{N'} \rightarrow \infty} \lim \left[ D_s= 1+\tfrac{\ln(L) - \ln(2)}{\ln(2{\cdot}{N'})}\right] 
\end{equation}
where $ D_h $ is again the Hausdorff--Besicovitch dimension,  $N' = N - 1$, and $ L $ is the curve length after embedding in the unit square. \textit{It is important to observe that $ D_s = D_h $ only at the limit when $ N' \rightarrow \infty $, for all other $ N' $ values, $ D_h > D_s $.} Thus $ D_s $s obtained with same $ N' $s are the only fit to be compared. $D_s$ uncertainty may be estimated through its variance \cite{Sevcik1998a}:
\begin{equation}\label{E:varD}
	\begin{split}
		\text{var}(D_s)&= \tfrac{N'}{L^2 \cdot \ln(2 \cdot N')^2} \sum_{i=1}^{N}\dfrac{(\Delta y_i - \overline{\Delta y})^2}{N'} = s^2\left( D_s\right)\\
		& \implies  s\left( D_s\right)  = \sqrt{\tfrac{N'}{L^2 \cdot \ln(2 \cdot N')^2} \sum_{i=1}^{N}\dfrac{(\Delta y_i - \overline{\Delta y})^2}{N'}}
	\end{split}
\end{equation}
where $ \Delta y_i $ is segment $ i $ length, distance between points  $ (x_{i-1},y_{i-1}) $ and $ (x_i,y_i) $ of the transform and $ \overline{\Delta y} $ is those segments\textquoteright{} mean. $D_s$ means and variances may be used to compare empiric $D_s$ values \cite[Sections 5.2.1 and 5.2.4]{DSuze2010a}\cite{Sevcik2018} based on Vysochanskyj--Petunin inequality \cite{Vysochanskij1980, Vysochanskij1983}.

\subsection{Random sequences generation for calibration purposes}\label{S:FractCalibr}

As indicated in Subsection \ref{S:TeorAnFrac} to use them as reference in this study, Monte Carlo simulation was used to generate$ M= 100 $ traces of $ N=1329 $ points each of white noise (see Subsection \ref{S:RandFract}) of \textit{white noise}. White noises simulated were sequences of Gaussian variables of type $ N(0,1) $ [Eq. (\ref{E:BoxMuller})]. Estimated fractal dimension $ D_s $ was calculated for each simulated trace as indicated in Subsection \ref{S:IntFracAnal}, $ D_s $ [Eq. (\ref{E:D_h})] and $ \text{var}(D_s) $ [Eq. (\ref{E:varD})]. Equation  (\ref{E:varD}) intra sequence $ D_s $ variance, due to unit square--embedded waveform $ \Delta y_i $ in each $ j $th sequence. Each sequence is a (short?) segment from an infinite sequence with same properties. Every sequence of length  $ N < \infty $ generated as indicated may be used to produce a different approximation to $ D_h $ depending on where is the sample taken. To estimate $ D_s $ and its uncertainty means were calculated for $ M $ simulated traces as 
\begin{equation}\label{E:DmedWithin}
	\bar{D}_{s}=\sum_{j=1}^{M}\dfrac{D_{s,j}}{M}.
\end{equation}
The variance od $ D_s  $\textit{\textbf{between}} curves (which may be visualised as random sets  $ N $ points sampled from an infinitely curve) for the $ M $ waveforms sets was determined as
\begin{equation}\label{E:VarWithin}
	\text{var}_b(\bar{D}_{s}) = \sum_{j=1}^{M}\dfrac{(D_{s,j}-\bar{D}_{s,w})^2}{M-1}.
\end{equation}
Then, total variance, $ \text{var}_t(\bar{D}_{s,w}) $, was determined as
\begin{equation}\label{E:varTotal}
	\text{var}_t(\bar{D}_{s,w}) =\text{var}_b(\bar{D}_{s}) + \left[ \overline{\text{var}_w(D_{s})}=\sum_{j=1}^{M}\tfrac{\text{var}(D_{s,j})}{M}\right] ,
\end{equation}
where $ \overline{\text{var}_w(D_{s})} $  is the average variance within traces and $ \text{var}(D_{s,j}) $ calculated with Eq. (\ref{E:varD}).

$ \bar{D}_{s,b} $ and $ \text{var}_t(\bar{D_s}_)  $ were used to compare the difference in $ D_s $ from the empiric sequences  with Monte Carlo simulated sequences with known properties. In this manner it was possible to decide if empiric sequences were white, brow , or other kind noise, spite of  $N<\infty$ \cite{Sevcik1998a, Sevcik2018}. Statistical significance of the differences was established using the Vysochanskij and Petunin inequality (V-P, details in \ref{S:VysPet}). V-P inequality allows comparing means of samples coming from a population with \textit{unknown pdf as long as it is unimodal, and has a definite variance }\cite{Vysochanskij1979, Vysochanskij1980, Vysochanskij1983}. Due to the unknown pdf, the comparison using V-P inequality tends to declare as not different to means that are weakly statistically significantly distinct  (Statistical error of type II \cite{Wilks1962}), but is $ \left(1- \tfrac{4}{9}\right)  \cdot 100 \approx  56$\% less likelier to produce an error of type II than classical Tchebichev inequality \cite{Tchebichef1867}, of which V-P inequality is a particular case. This means that parameter differences declared significant with the V-P inequality, are more significant than they appear. To compare the statistical significance of a difference between two $ D_s $ means  \cite[Adaptwd from Eqs. (34) and (35)]{DSuze2010a} it is required only that 
\begin{equation}
	P \left( \left|  \bar{D}_{s,1}-\bar{D}_{s,2} \right| = 0 \right) = P \left( \left| {\Delta \bar{D_s}_{(1,2)}} \right| = 0 \right)  \leq  \frac{2}{9\sqrt{\text{var}_t\left( \bar{D}_{s,1}\right)+\text{var}_t\left( \bar{D}_{s,2}\right)}}
\end{equation}
where the subindex t indicates total variance, and
\begin{equation}\label{E:AlfaVyso}
	P\left[\tfrac{\Delta \bar{D_s}_{1,2}}{\sqrt{\text{var}_t\left( \bar{D}_{s_{1}}\right)+\text{var}_t\left( \bar{D}_{s,2}\right)} }  \geq \left( \xi=\sqrt{\tfrac{40}{9}} = 2.108 \ldots \right) \right]  \leq 0.05 = \alpha
\end{equation}
where $ \alpha $ is the largest probability we accept to declare a difference statistically significant. If $ D $\textquoteright{}s pdf is known and Gaussian Eq. (\ref{E:AlfaVyso}) constant would not be  $ \xi=\sqrt{\tfrac{40}{9}}=2.108 \ldots$ as specified by the V-P inequality, but $ \xi=1.959 \ldots $.  This is true for any case where the pdf is known, thus we should write Eq. (\ref{E:AlfaVyso}) somewhat like $ \xi \lessapprox \sqrt{\tfrac{40}{9}}$ within the parentheses, but this is unnecessary due to the demonstrated inequality \cite{Vysochanskij1979, Vysochanskij1980, Vysochanskij1983}.

Eqs. (\ref{E:DmedWithin}) to (\ref{E:varTotal}) were also calculated for sets of $ M=100 $ Brownian sequences with $ N=1329 $ like
\begin{equation}
	b(t+\Delta t)=b(t) + N(0,\;1)
\end{equation}
and were used as standards to calculate the probability of significance between their estimated  $ \bar{D}_s $ versus $ D_s $ calculates for and empirical distribution of interest. Their fractal dimension $ D_h $, when $ N \rightarrow \infty $ is $ 1.5 $. An example may be found in \cite{Sevcik2018}. The $ N[0,1] $ variable was generated as indicated in Eq. (\ref{E:BoxMuller}).

\subsection{Generating uniform (rectangular) random variables}\label{S:Uniform}

Uniform random variables of  $ U[0,1] $ type were generated. Fundamental to all Monte Carlo simulations \cite{Dahlquist1974} is a good uniform (pseudo) random (PRNG) number generator. Data for all numerical simulations carried out in this work were produced using random numbers ($ r $) with continuous rectangular (uniform) distribution in the closed interval $ [0,1] $ or $ U [0, 1] $. All $ U [0, 1] $ [Eq. (\ref{E:U01})] were generated using the 2002/2/10 initialization-improved 623-dimensionally evenly distributed \cite{Equidist2020} uniform pseudo random number generator MT19937 algorithm \cite{Matsumoto1998, Matsumoto2000}. The procedure used makes exceedingly unlikely ($ P = 2^{−64} \approx  5.4 \cdot 10^{−20} $ ) that the same sequence, $ \{r_i \} $, of $ U [0, 1] $ is used twice. Calculations were programmed in C++ using g++ version 5.4.0 20160609 with C++14 standards, under Ubuntu GNU Linux version 18.4. For further details on random number generator initializing see Sevcik \cite{Sevcik2018}.

Rectangular random variables  fulfil the following property  
\begin{equation}\label{E:U01}
	U[a,b]=
	\begin{cases}
		0 \quad \text{if} \quad x < a \\
		k = \frac{1}{b-a}\quad \text{if} \quad a \leqslant  x  \leqslant b \\
		0 \quad \text{if} \quad  x > b
	\end{cases}
\end{equation}  
with mean $ \E(x)=\frac{b-a}{2} $, $ x $, $ k $  may be a real number  \cite{Sevcik1998a} or an integer \cite{Sevcik1998a, Sevcik2018}. 

\subsection{Generating random normal deviates}\label{S:Normal}

Random normal variables of the ding $ N[\mu, \sigma]=N[0,1] $ were generated with the \citet{Box1958}\cite{BoxMuller202} algorithm
\begin{equation}\label{E:BoxMuller}
	N[0,1]=
	\begin{cases}
		\sin[2 \pi r_1] \sqrt{-2 \ln[r_2]} \\
		\\
		\cos[2 \pi r_1] \sqrt{-2 \ln[r_2]}
	\end{cases}
\end{equation}
where $ r_1 $ and $ r_2 $ are two uniformly distributed random variates of the kind $ U[0,1.] $.

\subsection{On Cauchy distributed variables}\label{S:Cauchy}
The Cauchy distribution is symmetric about its median (location factor, $ \hat{\mu} $) and
has a width (dispersion) factor ($ \lambda $) and its pdf is
\begin{equation}\label{E:Cauchy_pdf}
c(x) =\frac{1}{\pi}\cdot\frac{\lambda}{\lambda^2+\left(x-\hat{\mu} \right)² }
\end{equation}
and also has a PDF like
\begin{equation}\label{E:Cauchy_PDF}
C(x) =\frac{1}{\pi}\cdot  \arctan \left( \frac{x-\hat{\mu}}{\lambda}\right)  +\frac{1}{2}
\end{equation}
where \textit{arctan} refers to the \textit{arc tangent} trigonometric function. Cauchy pdf has no central moments \cite{Wilks1962, Walck1996}, and thus has no defined mean, variance, skewness or kur-tosis \citep{Barnett1966, Walck1996, Lohninger2012}. If conventional equations for sample $ \bar{y} $ or $ s^2 (y) $ values are calculated, they will be seen to vary wildly and wild variability increases with sample size \citep{Runkel2012}. Cauchy–type data can be described and compared only with nonparametric statistics. From Eq. (\ref{E:Cauchy_PDF}) it is possible to prove that a 95\% CI of a Cauchyan variable is equivalent to $ \approx \mu \pm 12.706\lambda $ which gives a $ \approx 25.412 \lambda $ span to the 95\% CI. A Cauchy random variable is prone to assume values very far away from its median, outliers are very commonly observed as sample size increases \citep{Barnett1966, Walck1996, Lohninger2012}.

\subsection{The Vysochanskij-Petunin Inequality \cite{Vysochanskij1980, Vysochanskij1983}.}\label{S:VysPet}

\begin{theorem}[Vysochanskij-Petunin]:\label{Te:VysoPet}
	Let X be a random variable with unimodal distribution, mean ${\mu}$ and finite, non-zero variance ${\sigma^2}$. Then, for any ${\lambda > \sqrt{\tfrac{8}{3}} = 1.63299}\ldots$
	\begin{equation}\label{E:VysoPet}
		P(\left| X- \mu \right| \geq \lambda \sigma) \leq \tfrac{4}{9\lambda^2}=\epsilon.
	\end{equation}
\end{theorem}

\subsubsection{Using the inequality to calculate significance when $ \lambda > \sqrt{\tfrac{8}{3}}$.}\label{S:VPT_Holds}
Then Theorem \ref{Te:VysoPet} holds even with heavily skewed distributions and puts bounds on how much of the data is, or is not \comillas{in the middle}. Setting ${\epsilon = 0.05}$ then ${\lambda=\pm \left ( \sqrt{\tfrac{80}{9}}\approx2.981\right )}$, for a two tailed test. By virtue of eq.~(\ref{E:VysoPet}) no matter which unimodal distribution, no matter how skewed, there will be <2.5\% chance that a datum will belong to a population with mean $\mu$ and variance $\sigma^2$ if it lays farther than $\pm2.981\sigma$ from $\mu$. Please note that if the probability distribution function of data is Gaussian $\pm1.96\sigma$ suffices to reach the same confidence level. We introduced the $ \epsilon $ variable in Eq. (\ref{E:VysoPet}) for the statistical argumentation that follows.

Let $\bar{D}_{s,1}$ and  $s^2(D_{s,1})$ be the mean and variance estimated for the random variable ${D_{s,1}}$, and $\bar{D}_{s,2}$ and  $s^2(D_{s,2})$ be the mean and variance estimated for the random variable ${D_{s,2}}$ (linearly independent from $D_{s,1}$), then
\begin{equation}
	\Delta \bar{D}_{s,(1,2)}=\bar{D}_{s,1}-\bar{D}_{s,2}
\end{equation}
and the estimated variance of $\Delta\bar{D}_{s,(1,2)}$ is
\begin{equation}
	s^2(\Delta{\bar{D}_{s,(1,2)}})=s^2(D_{s,1})+s^2(D_{s,2}).
\end{equation}
By virtue of the V-P inequality [Eq. (\ref{E:VysoPet})]
\begin{equation}\label{E:VysoPetVP}
	P \left(\tfrac{ \left|\Delta{\bar{D}_{s,(1,2)}} \right|}{s(\Delta{\bar{D}_{s,(1,2)}})} \geqslant \sqrt{\tfrac{4}{9 \lambda^2}}\right )  = P\left(\tfrac{2\left|\Delta{\bar{D_s}_{1,2}} \right|}{3s(\Delta{\bar{D}_{s,(1,2)}})} \leqslant  \lambda\right ) \leqslant \epsilon
\end{equation}
which means Ec. (\ref{E:VysoPet}) that
\begin{equation}\label{E:P_VP}
	\epsilon=\frac{2}{3 \left[ \frac{\Delta\bar{D}_{s,(1,2)}}{s(\Delta{\bar{D}_{s,(1,2)}})} \right ]}.
\end{equation}

Therefore, there is $\leq \epsilon$ probability that $\left|\Delta{\bar{D}_{s,(1,2)}} \right|\neq0$ due to random sampling variation, and $ \Delta{\bar{D} }_{s,(1,2)} \neq 0 $ with a confidence level $P\leq\epsilon$. Thus:
\begin{equation}\label{E:P_VP2}
	P \left[\tfrac{ \left|\Delta{\bar{D}_{s,(1,2)}} \right|}{s(\Delta{\bar{D}}_{s,(1,2)})} \geq \left ( \sqrt{\tfrac{40}{9}}\approx2.108\right )\right ] \leq \epsilon
\end{equation}

\noindent{for a test with only one tail, since in this case we are only interested with one alternative, that $\left|\Delta{\bar{D}_{s,(1,2)}} \right|\neq0$. Again, when the probability distribution function is Gaussian, the value of $\lambda$ for the one tailed case, would be $\approx1.65$ instead of the $\approx2.108$ demanded by equation~(\ref{E:VysoPet}).}

\subsubsection{Using the inequality to calculate significance when $ \lambda \leqslant \sqrt{\tfrac{8}{3}}$.}\label{S:VPT_NotHolds}

Theorem \ref{Te:VysoPet} does not hold, but you can still get an answer  using Eq. (\ref{E:VysoPet}) as shown in the following Corollary \ref{Co:AtLeast} of the V-P inequality.

\begin{corollary}:\label{Co:AtLeast}
	If $ \lambda \leqslant \sqrt{\tfrac{8}{3}}$ then $ P $ is not known but is bound as
	\begin{equation}\label{E:AtLeast}
		\tfrac{1}{6} \leqslant P \leqslant 1
	\end{equation}
	\begin{proof}
		As indicated by Eq. (\ref{E:VysoPet}), $\epsilon= \tfrac{4}{9\lambda^2} $ then if
		\begin{equation}\label{E:VysoPet_max}
			\lambda \leqslant  \sqrt{\tfrac{8}{3}}  \implies  
			\epsilon  \geqslant \tfrac{4\cdot3 }{9\cdot 8} \geqslant \tfrac{1}{6} = \epsilon_{th}
		\end{equation}
		then $ \epsilon_{th} $ is the value of $ \epsilon $ at the threshold required for Theorem \ref{Te:VysoPet} to hold. When $ \lambda \leqslant \sqrt{\tfrac{8}{3}}$ the predicted value of $ \epsilon \geqslant \epsilon_{th} $, and since there is no limit to this inequality, it may happen that the estimated $ \epsilon > 1 $, but there is no probability $ P >  1 $, by definition. Thus when $ \lambda \leqslant \sqrt{\tfrac{8}{3}}$ it follows that  $ \epsilon $ does not express a probability.
		But since if $ \lambda \leqslant \sqrt{\tfrac{8}{3}}$ Eq. (\ref{E:VysoPet_max}) gives the highest $ P $ value which may be estimated with the V-P inequality is $P= \frac{1}{6}$, $ P $ is undetermined but  bound in the interval:	
		\begin{equation}
			\then  \left (\lambda \leqslant \sqrt{\tfrac{3}{8}}\right) \implies \left( \tfrac{1}{6} \leqslant P \leqslant 1\right) . 
		\end{equation}
	\end{proof}
\end{corollary}

\end{appendices}

\section*{Acknowledgements.}

All expenses were covered by Hospital La Fuenfr\'{\i}a, Servicio Madrile\~{n}o de salud (SERMAS). An non arbitered preliminary version of this article may be found in arXiv as \url{http://export.arxiv.org/pdf/2011.09514}. The authors thank Prof. Rafael Apitz for its critical reading of the manuscript an his comments which helped to increase its clarity.

This manuscript was written in \LaTeX{ }using \TeX{}studio  for Ubuntu  GNU Linux{}, \url{http://www.texstudio.org}), an open source free  \LaTeX{ }editor with \TeX{} Live 20 (\url{https://www.tug.org/texlive/}).  Most graphics were built using \textbf{Libr}eOffice Calc v. 7.1.2.2 and combined into figures using the GNU Image Manipulation Program v. 2.10.24 (GIMP). All these programs, free  and open sourced, are available for Apple{} OS-X, \textit{MicroSoft} Windows{} and GNU Linux.


\end{document}